\documentclass[twocolumn,showpacs,preprintnumbers,amsmath,amssymb]{revtex4}
\usepackage{graphicx}
\usepackage{dcolumn}
\usepackage{bm}

\newcommand{\hH}{\hat{H}}
\newcommand{\ha}{\hat{a}}
\newcommand{\hap}{\hat{a}^+}

\begin{document}
\title{Zitterbewegung (trembling motion) of electrons in semiconductors: \\
             a Review}
\date{\today}
\author{Wlodek Zawadzki*}
\author{Tomasz M. Rusin\dag}
\email{zawad@ifpan.edu.pl}

\affiliation{*Institute of Physics, Polish Academy of Sciences, Al. Lotnik\'ow 32/46, 02-688 Warsaw, Poland\\
           \dag Orange Customer Service sp. z o. o., ul. Twarda 18, 00-105 Warsaw, Poland}

\begin{abstract}
We review recent research on Zitterbewegung (ZB, trembling motion) of electrons in semiconductors.
A brief history of the subject is presented, the trembling motion in semirelativistic and spin systems
is considered and its main features are emphasized. Zitterbewegung of charge carriers in monolayer and
bilayer graphene as well as in carbon nanotubes is elaborated in some detail. We describe effects
of an external magnetic field on ZB using monolayer graphene as an example. Nature of electron ZB
in crystalline solids is explained. We also review various simulations of the trembling motion
in a vacuum and in semiconductors, and mention ZB-like wave phenomena in sonic and photonic periodic
structures. An attempt is made to quote all the relevant literature on the subject.
\end{abstract}


\maketitle
\section{Introduction and brief history}

This review article is concerned with a somewhat mysterious phenomenon known in the literature
under the German name of ``Zitterbewegung'' (trembling motion). Both the phenomenon and its name
were conceived by Erwin Schrodinger who, in 1930, published the paper
{\it Ueber die kraeftefreie Bewegung in der relativistischen Quantenmechanik}
in which he observed that in the Dirac equation, describing relativistic electrons in a vacuum,
the $4\times 4$ operators corresponding to components of relativistic velocity do not commute
with the free-electron Hamiltonian~\cite{Schroedinger1930}.
In consequence, the electron velocity is not a constant of the
motion also in absence of external fields. Such an effect must be of a quantum nature as it does not
obey Newton's first law of classical motion. Schrodinger calculated the resulting time dependence of
the electron velocity and position concluding that, in addition to classical motion,
they experience very fast periodic oscillations
which he called Zitterbewegung (ZB). Schrodinger's idea stimulated numerous theoretical investigations
but no experiments since the predicted frequency $\hbar\omega_Z\simeq 2m_0c^2\simeq 1$ MeV and
the amplitude of about $\lambda_c=\hbar/mc\simeq 3.86\times 10^{-3}$\AA\ are not accessible to
current experimental techniques.
Huang~\cite{Huang1952} put the theory on a more physical basis calculating
averages of velocity and position operators.
It was recognized that the ZB is due to an interference of states corresponding to the positive and
negative electron energies resulting from the Dirac equation~\cite{BjorkenBook,SakuraiBook,GreinerBook}.
Lock~\cite{Lock1979} showed that, if an electron is represented by a wave packet,
its ZB has a transient character, i.e. it disappears with time.

It was conceived years later that the trembling electron motion should occur also in
crystalline solids if their band structure could be represented
by a two-band model reminiscent of the Dirac equation.
The first paper in this line, published in 1970 by Lurie and Cremer~\cite{Cremer1970},
was concerned with superconductors, in which
the energy-wave vector dependence is similar to the relativistic relation. Similar
approach was applied to semiconductors twenty years later using a model of two energy bands
\cite{Cannata1990,Ferrari1990,Vonsovskii1990,ZawadzkiHMF}.
However, an intense interest in ZB of electrons in semiconductors was launched only in 2005.
Zawadzki~\cite{Zawadzki2005KP} used a close analogy between the $\bf k \cdot \bf p$ theory
of energy bands in narrow gap semiconductors (NGS)
and the Dirac equation for relativistic electrons in a vacuum to show that one should
deal with the electron ZB in NGS which would have much more favorable frequency and
amplitude characteristics than those in a vacuum. On the other hand,
Schliemann {\it et al.}~\cite{Schliemann2005} demonstrated
that the spin splitting of energies linear in~$k$, caused by the inversion asymmetry in
semiconductor systems (the Bychkov-Rashba splitting),
also leads to a ZB-type of motion if the electron wave packet has a non-vanishing initial momentum.
The above contributions triggered a wave of theoretical considerations for various
semiconductor and other systems. It was recognized that the phenomenon of ZB occurs every
time one deals with two or more interacting energy bands~\cite{Cserti2006,Winkler2007,Rusin2007a}.

It was shown that, indeed, when the electron is represented by a wave packet, the ZB has a transient
character~\cite{Rusin2007b}. Considering graphene in a magnetic field it was
demonstrated that, if the electron spectrum is discrete, ZB contains many frequencies and it
is sustained in time~\cite{Rusin2008}. It was pointed out that the trembling electrons should emit
electromagnetic radiation if they are not in their eigenstates~\cite{Rusin2009}. The physical
origin of ZB was analyzed and it turned out that, at least in its ``classical'' solid state version
analogous to the ZB in a vacuum, the trembling motion represents simply oscillations of electron
velocity due to the energy conservation as the particle moves in a periodic potential~\cite{Zawadzki2010}.

As mentioned above, in a vacuum the ZB characteristics
are not favorable. In solids, the ZB characteristics are much better but it is
difficult to observe the motion of a single electron.
However, recently Gerritsma {\it et al.}~\cite{Gerritsma2010} simulated
experimentally the Dirac equation
and the resulting electron Zitterbewegung with the use of trapped ions and laser excitations.
The power of the simulation method is that one can adjust experimentally the essential
parameters of the Dirac equation:~$mc^2$ and~$c$, and thus achieve more favorable values of
the ZB frequency and amplitude. The experimental results obtained by Gerritsma {\it et al.}
agreed well with the predictions of Zawadzki and Rusin~\cite{Zawadzki2010}.
Interestingly, it turned out that analogues of ZB can occur also in classical wave propagation
phenomena. Several predictions were made, but in two systems, namely macroscopic sonic crystals
\cite{ZhangLiu2008}, and photonic superlattices~\cite{Dreisow2010}, the ZB-like effects were actually observed.
Finally, there has been growing recognition that the mechanisms responsible for ZB in solids are related
to their other properties, for example to the electric conductivity.

Thus the subject of our interest is not only quickly developing but also quite universal.
From an obscure, perplexing and somewhat marginal effect that would probably never be observed,
the Zitterbewegung has grown into a universal, almost ubiquitous phenomenon that was experimentally
simulated in its quantum form and directly observed in its classical version.
Our article summarizes the first five years of the intensive development which can be
characterized as the ``Sturm und Drang'' period, to use another pertinent German term.
We concentrate mostly on the ZB in semiconductors but mention other systems, in particular different
ZB simulations by trapped ions and atoms, as these seem to be most promising for future experimental
observations. We also briefly review important papers describing the ZB of free relativistic electrons
in a vacuum since they inspired early considerations concerning solids. The subject of Zitterbewegung
has been until now almost exclusively theoretical. Below we quote mostly derivations and figures from
our own papers, not because we believe that they are
the only important ones, but because of the copyright restrictions.

The review is organized in the following way. In Sec. II we present descriptions of ZB for
semi-relativistic, spin, and nearly-free electron Hamiltonians and quote papers on other model systems.
Section III treats the trembling motion in bilayer graphene, monolayer graphene and carbon nanotubes.
In Sec. IV we consider the ZB in the presence of an external magnetic field and quote related works.
Section V is concerned with the origin of ZB in crystalline solids. There follows short section VI
in which we mention work relating ZB to calculations of electric conductivity. In Sec. VII we describe
papers on the ZB of free relativistic electrons, necessary to understand the trembling motion in
solids. Section VIII contains a very brief introduction to simulations of the Dirac equation and
the resulting ZB in absence of fields and in a magnetic field. The section is completed by
summaries of related papers. In Sec. IX we describe wave ZB-like effects in non-quantum periodic systems.
The review is terminated by discussion and conclusions.

\section{ZB in Model Systems}

We begin our considerations of electron ZB in semiconductors by using the so called
relativistic analogy~\cite{Zawadzki2005KP}. This way we can follow simultaneously the procedure of Schrodinger
and derive corresponding relations for narrow gap semiconductors.
It was noted in the past that the~$E(k)$ relation between the energy~$E$ and the wave vector~$k$ for
electrons in NGS is analogous to that for relativistic electrons
in a  vacuum~\cite{Zawadzki1966,ZawadzkiOPS,ZawadzkiHMF}.
The semi-relativistic phenomena appear at electron velocities of $v\simeq 10^7-10^8$ cm/s, much lower than
the light velocity~$c$. The reason is that the maximum velocity~$u$ in semiconductors, which plays the role
of~$c$ in a vacuum, is about $10^8$ cm/s. To be more specific, we use the $\bf k \cdot \bf p$ approach to
InSb-type semiconductors~\cite{Kane1957}. Taking the limit of of large spin-orbit energy, the resulting
dispersion relation for the conduction and the light-hole bands is~$E=\pm E_p$, where
\begin{equation} \label{E_2bands}
 E_p = \left [\left(\frac{E_g}{2}\right)^2 + E_g \frac{p^2}{2m_0^*} \right]^{1/2}.
\end{equation}
Here~$E_g$ is the energy gap and~$m_0^*$ is the effective mass at the band edge.
This expression is identical to the relativistic relation for electrons in a vacuum,
with the correspondence $2m_0c^2\rightarrow E_g $ and $m_0 \rightarrow m_0^*$.
The electron velocity~$\bm v$ in the conduction band described by~(\ref{E_2bands}) reaches a saturation
value as~$p$ increases. This can be seen directly by calculating $v_i=\partial E_p /\partial p_i$
and taking the limit of large~$p_i$, or by using the analogy
$c=(2m_0c^2/2m_0)^{1/2}\rightarrow (E_g/2m_0^*)^{1/2}=u$. Taking the experimental
parameters~$E_g$ and~$m_0^*$ one calculates a very similar
value of $u\simeq 1.3\times10^8$ cm/s for different semiconductor compounds.
Now we define an important quantity
\begin{equation} \label{E_LambdaZ} \lambda_Z = \frac{\hbar}{m_0^*u} \end{equation}
which we call the length of Zitterbewegung for reasons given below. We note that it corresponds to the
Compton wavelength $\lambda_c=\hbar/m_0c$ for electrons in a vacuum.

Next we consider the band Hamiltonian for NGS. It is
derived within the model including~$\Gamma_6$ (conduction),~$\Gamma_8$ (light and heavy hole),
and~$\Gamma_7$ (split-off) bands and it represents an $8\times 8$ operator matrix~\cite{Kane1957}.
We assume, as before, $\Delta \gg E_g$ and omit the free electron terms since they are
negligible for NGS. The resulting $6\times 6$ Hamiltonian has~$\pm E_g/2$ terms on
the diagonal and linear~$\hat{p}_i$ terms off the diagonal, just like in the Dirac equation
for free electrons. However, the three $6\times 6$ matrices~$\bm \hat{\alpha}_i$ multiplying the
momentum components~$\hat{p}_i$ do not have the properties of $4\times 4$ Dirac matrices, which
considerably complicates calculations. For this reason,
with only a slight loss of generality, we take $\hat{p}_z\neq 0$ and
$\hat{p}_x=\hat{p}_y=0$. In the~$\hat{\alpha}_3$ matrix, two rows and columns corresponding
to the heavy holes contain only zeros and they can
be omitted. The remaining Hamiltonian for the conduction
and the light hole bands reads
\begin{equation} \label{H_WZ}
 \hat{H} = u\hat{\alpha}_3\hat{p}_z + \frac{1}{2}E_g\hat{\beta},
\end{equation}
where~$\hat{\alpha}_3$ and~$\hat{\beta}$ are the well-known $4\times 4$ Dirac matrices~\cite{DiracBook}.
The Hamiltonian~(\ref{H_WZ}) has the form appearing in the Dirac
equation and in the following we can use the procedures of
relativistic quantum mechanics (RQM). The electron velocity is
$\dot{\hat{z}}=(1/i\hbar)[\hat{z},\hat{H}]=u\hat{\alpha}_3$. The eigenvalues
of~$\hat{\alpha}_3$ are~$\pm 1$, so that the eigenvalues of~$\dot{\hat{z}}$ are, paradoxically,~$\pm u$.
In order to determine~$\hat{\alpha}_3(t)$ we calculate~$\dot{\hat{\alpha}}_3(t)$
by commuting~$\hat{\alpha}_3(t)$ with~$\hat{H}$ and integrating the result with respect
to time. This gives~$\dot{\hat{z}}(t)$ and we calculate~$\hat{z}(t)$ integrating
again. The final result is
\begin{equation} \label{z(t)}
 \hat{z}(t)= \hat{z}(0) + \frac{u^2\hat{p}_z}{\hat{H}}t + \frac{i\hbar u}{2\hat{H}} \hat{A}_0
      \left[ \exp\left( \frac{-2i\hat{H}t}{\hbar} \right)-1 \right],
\end{equation}
where $\hat{A}_0=\hat{\alpha}_3(0)-u\hat{p}_z/\hat{H}$. There is $1/\hat{H}=\hat{H}/E_p^2$.
The first two terms of~(\ref{z(t)}) represent the classical electron motion.
The third term describes time dependent oscillations with the frequency of $\omega_Z \simeq E_g/\hbar$.
Since $\hat{A}_0\simeq 1$, the amplitude of oscillations is $\hbar u/2\hat{H} \simeq \hbar/2m_0^*u=\lambda_Z/2$.
In RQM the analogous oscillations are called Zitterbewegung,
which explains the name given above to~$\lambda_Z$.
The expression obtained by Schrodinger for ZB of free relativistic
electrons in a vacuum is identical to that given by~(\ref{z(t)}) with the use of the above relativistic analogy.
In RQM it is demonstrated that ZB is a result of interference between states of positive
and negative electron energies~\cite{BjorkenBook,GreinerBook,SakuraiBook}.
Clearly, one can say the same of ZB in semiconductors calculated according to the above model.
However, as we show below, the origin of ZB in crystalline solids can be interpreted in more
physical terms. The magnitude of~$\lambda_Z$ is essential.
There is $\lambda_z=\lambda_c(c/u)(m_0/m_0^*)\simeq 0.89(m_0/m_0^*)$\AA\ since, as mentioned
above, $u=\simeq 1.3 \times 10^8$ cm/s for various materials. We estimate:
for GaAs ($m_0^*\simeq 0.067m_0$) $\lambda_Z \simeq 13$\AA, for InAs ($m_0^*\simeq 0.024m_0$)
$\lambda_Z \simeq 37$\AA, for InSb ($m_0^*\simeq 0.014m_0$) $\lambda_Z \simeq 64$\AA.
Thus, in contrast to a vacuum, the length of ZB in semiconductors can be quite large.
However, one should bear in mind that the above derivations,
as well as the original procedure of Schrodinger's, use only operator considerations, whereas
physical observables are given by quantum averages. We show below that, if one calculates such averages
using electron wave packets, the amplitude of ZB may be considerably smaller than~$\lambda_Z$.

Next, we briefly consider another example of ZB proposed by Schliemann
{\it et al.}~\cite{Schliemann2005}. It is based on the so-called Bychkov-Rashba (BR) spin
splitting caused by structure inversion asymmetry in two-dimensional
semiconductor heterostructures~\cite{Bychkov1984}.
The interaction describing this splitting is
\begin{equation} \label{H_BR}
\hat{H}_{BR}= \frac{\alpha}{\hbar} (p_x\sigma_y - p_y\sigma_x),
\end{equation}
where~$\bm p$ is the momentum of an electron confined in two-dimensional geometry, and~$\bm \sigma$
is the vector of Pauli matrices. The coefficient~$\alpha$ is to be calculated using details
of the structure~\cite{Zawadzki2004}. One can easily solve the eigenenergy equation
and obtain the spin energies $E=\pm \alpha k$. The important difference with the
case considered above is that here at~$k=0$ there is $\Delta E=0$, i.e. there
is no gap. The complete Hamiltonian is $\hat{H}={\bm p}^2/2m^*+\hat{H}_{BR}$, where~$m^*$ is the effective mass.
The position operator in the Heisenberg picture has the standard form
\begin{equation}
\hat{\bm r}(t)= e^{i\hat{H}t/\hbar}\hat{\bm r}(0)e^{-i\hat{H}t/\hbar}.
\end{equation}
One calculates $\hat{\bm {r}}(t)$ explicitly using the Hamiltonian and averages it
employing a Gaussian wave packet of the width d centered at the wave
vector $k_{0x}=0$ and $k_{0y}\neq 0$. In case $dk_{0y}\gg 1$, Schliemann
{\it et al.} obtained
\begin{equation} \label{Sch_x(t)}
\langle \psi|\hat{\bm x}(t)|\psi\rangle = \frac{1}{2k_{0y}}
 \left[1-\cos\left(\frac{2\alpha k_{0y}t}{\hbar} \right)\right].
\end{equation}
The above result describes ZB with the frequency given by
$\hbar\omega_Z=2\alpha k_{0y}$, where $\hbar\omega_Z$ is the excitation energy
between the two branches of the Bychkov-Rashba energies at $k=k_{0y}$.
It is seen that ZB is absent for $k_{0y}=0$.
Similar results for ZB are obtained if, instead of the spin splitting
due to structure inversion asymmetry, one uses a two-dimensional version
of the spin splitting due to crystal inversion asymmetry (Dresselhaus
splitting~\cite{Dresselhaus1955}) described by the Hamiltonian
\begin{equation}
\hat{H}_{D}= \frac{\beta}{\hbar} (p_x\sigma_x - p_y\sigma_y).
\end{equation}

To demonstrate universality of ZB in the two-band situation we consider the well-known case
of nearly free electrons in which the periodic lattice potential
$V({\bm r})$ is treated as a perturbation (see~\cite{Rusin2007a}).
Near the Brillouin zone boundary the Hamiltonian
has, to a good approximation, a $2\times 2$ form (spin is omitted)
\begin{equation} \label{defhH}
 \hat{H} = \left(\begin{array}{cc} E_{\bm k+\bm q} & V_{\bm q} \\
                    V_{\bm q}^* & E_{\bm k} \\ \end{array}\right),
\end{equation}
where $V^*_{\bm q}=V_{-\bm q}$ are the Fourier coefficients in the expansion of $V({\bm r})$,
and $E_{\bm k}=\hbar^2 k_z^2/2m_0$ is the free electron energy. The $2\times 2$ quantum
velocity~$\hat{v}_z$ can now be calculated and the acceleration~$\dot{\hat{v}}_z$ is computed
in the standard way. Finally, one calculates the displacement matrix~$\hat{z}_{ij}$.

\begin{figure}
\includegraphics[width=8cm,height=8cm]{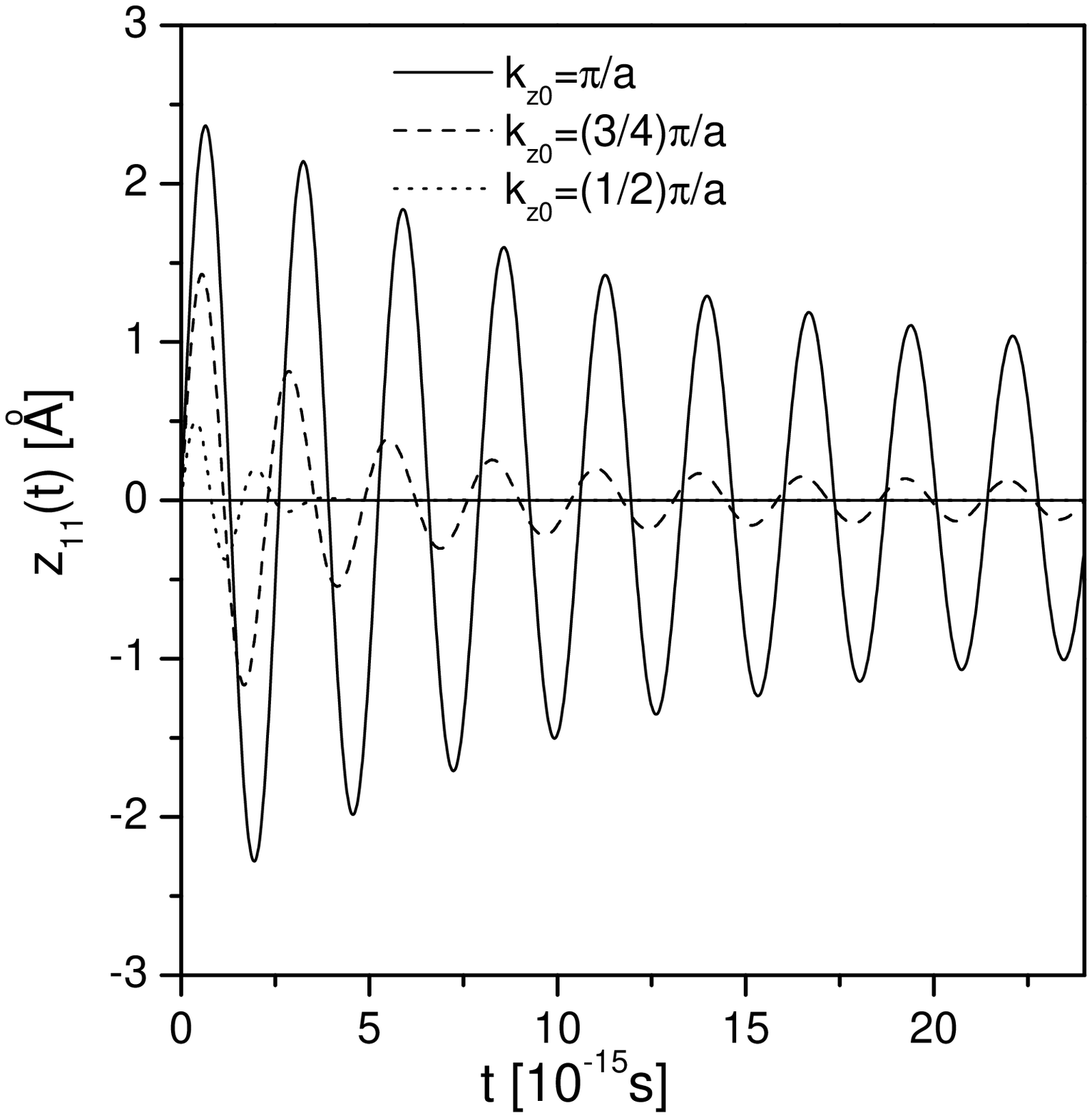}
\caption{\label{NGapFig2} Transient Zitterbewegung oscillations of nearly-free electrons  {\it versus} time,
          calculated for a very narrow wave packet centered at various~$k_{z0}$ values.
          The band parameters correspond to GaAs. After~\cite{Zawadzki2008}.}
\end{figure}

Since the ZB is by its nature not a stationary state but a dynamical phenomenon, it is natural
to study it with the use of wave packets. These have become a practical instrument
when femtosecond pulse technology emerged. Thus, in a more realistic
picture the electrons are described by wave packets
\begin{eqnarray}\label{defPacket}
 \psi(z) &=& \frac{1}{\sqrt{2\pi}}\frac{d^{1/2}} {\pi^{1/4}}
           \int_{\infty}^{\infty}\exp\left(-\frac{1}{2}d^2(k_z-k_{z0})^2\right) \times \nonumber \\
         && \exp(i{k_zz}) dk_z \left( \begin{array}{c} 1 \\ 0 \end{array} \right).
\end{eqnarray}
The electron displacement is calculated as an average of the position operator~$\hat{z}$ over the
above wave packet, see Figure~\ref{NGapFig2} and Reference~\cite{Zawadzki2008}.
The essential result is that, in agreement with Lock general predictions~\cite{Lock1979},
the ZB oscillations of the average electron position have a {\it transient} character, i.e.
they disappear with time on a femtosecond scale. The frequency of oscillations is $\omega_Z=E_g/\hbar$,
where $E_g=2|V_{\bm q }|$.

Lurie and Cremer~\cite{Cremer1970} in their very early paper described ZB of electrons in
superconductors using the fact that a Bogoliubov quasi-particle is described by a $2\times 2$ Hamiltonian
reminiscent of the Dirac equation with finite gap.
The resulting ZB has a frequency corresponding to the gap and a large amplitude.
However, this treatment was carried on the operator level without calculating physical averages.

Cannata {\it et al.}~\cite{Cannata1990} in an early paper described a one-dimensional periodic
chain using the tight-binding LCAO scheme and observed that the resulting two-band model bears
strong similarities to the Dirac equation (without spin). Using the relativistic analogy
this paper defined the effective ``rest mass'', the effective ``Compton wavelength'' and
predicts the Zitterbewegung of non-relativistic carriers with the amplitude of
about a lattice constant.

Shmueli {\it et al.}~\cite{Shmueli1995} considered tunnelling across a~$p-n$ tunnelling
diode with the motion on both
sides constrained by quantum wells. The dispersion curves~$E({\bm k})$ on~$n$ and~$p$ sides of the
diode resemble that of the relativistic DE. It was shown that in such a system an electron will
oscillate between the two QWs in a motion similar to ZB having the frequency proportional to
the tunnelling amplitude.

Jiang {\it et al.}~\cite{Jiang2005} studied numerically the dynamics of holes in degenerate~$\Gamma_8$ bands
described by the Luttinger Hamiltonian in the presence of a constant electric field. It was found
that the time-dependence of hole trajectories and their spin contain rapid oscillations reminiscent of
the Zitterbewegung. Frequencies of the oscillations are given by the differences of
light- and heavy-holes energies: $\hbar\omega(t)=E_L(t)-E_H(t)$, and they increase in time as the holes
are accelerated to higher~$k$ values by the electric field.

Bernardez {\it et. al}~\cite{Bernardes2006} described the ZB in spin space caused by a new kind
of spin-orbit interaction resulting from an inter-subband coupling in symmetric quantum wells.
In this case the ZB is characterized by cycloidal electron trajectories.

Winkler {\it et al.}~\cite{Winkler2007} considered the oscillatory dynamics of Heisenberg observables
such as position ${\bm r}(t)$, velocity ${\bm v}(t)$, orbital angular momentum ${\bm L}(t)$ and
spin ${\bm S}(t)$ in a variety of different systems described by the Bychkov-Rashba~\cite{Bychkov1984},
Luttinger~\cite{Luttinger1956} and Kane~\cite{Kane1957} Hamiltonians.
They illustrated similarities between their time
evolutions with the resulting ZB-like effects.

Demikhovskii {\it et al.}~\cite{Demikhovskii2010} described 3D hole system
having the effective spin~$3/2$ with the use of wave packets. For~$dk_0\gg 1$,
where~$k_0$ is the average packet wave vector and~$d$ is the packet widths, the initial wave packet
splits into two parts and the packet's center experiences the transient ZB. It was also shown that the
average angular momentum and spin vector undergoes a transient precession due to the interference of
the light- and heavy-hole states.

Cserti and David~\cite{Cserti2006} observed that the Hamiltonians mentioned above
and describing the Bychkov-Rashba and the Dresselhaus spin splitting,
monolayer and bilayer grapheme, nearly-free electrons, electrons in
superconductors, etc., can be represented in the general form
\begin{equation} \label{H_Cserti}
\hat{H} = \epsilon(\bm p) \hat{\bm 1} + {\bm \Omega}^T \hat{\bm S},
\end{equation}
where the one-particle dispersion is described by $\epsilon(\bm p)$ and the
second term has the form of an effective magnetic field $\Omega(\bm p)$ coupled
to the spin~$\hat {\bm S}$. Here~$T$ stands for the transpose of a vector, while~$\hat{\bm 1}$ is the
unit vector in spin space. One can use the Hamiltonian~(\ref{H_Cserti}) to calculate
time-dependent position operator in the Heisenberg picture and show
that it consists in general of the ``classical'' (mean) part and the ZB
part. For the Bychkov-Rashba coupling, see~(\ref{H_Cserti}), and for monolayer
graphene, see below, the ZB can be interpreted as a consequence of conservation of the
total angular momentum $J_z=L_z+S_z$, where ${\bm L}={\bm r} \times {\bm p}$
is the orbital angular momentum. This is in analogy to the results discussed in section~V,
where ZB is shown to be a consequence of the energy conservation.

David and Cserti~\cite{David2010} considered a general multi-band Hamiltonian and showed that in this case
one deals with a trembling motion which is a superposition of trembling motions corresponding to
all possible differences of energy eigenvalues. It was also shown, following
remarks of~\cite{Clark2008,Englman2008}, that the ZB amplitudes in the position operator are related to the
Berry connection matrix appearing in the expression of the Berry phase. We may add that a good example
of multi-frequency ZB motion is given by carriers in graphene in a magnetic field, (see~\cite{Rusin2008},
and Figure~\ref{GraphHFig1}), where the role of different eigenenergies is represent by different Landau levels.

Wilamowski {\it et al.}~\cite{Wilamowski2010} investigated microwave absorbtion in asymmetric Si quantum
wells in an external magnetic field and detected a spin-dependent component of the Joule heating
at the spin resonance. The observation was explained in terms of the Bychkov-Rashba spin-splitting
due to the structure inversion asymmetry with the resulting current-induced spin precession and
the ZB at the Larmor frequency.

\section{ZB in graphene}
Now we study in some detail the Zitterbewegung of mobile charge carriers in three modern materials:
bilayer graphene, monolayer graphene, and carbon nanotubes~\cite{Rusin2007b}.

\subsection{Bilayer graphene}
\begin{figure}
\includegraphics[width=8.5cm,height=8.5cm]{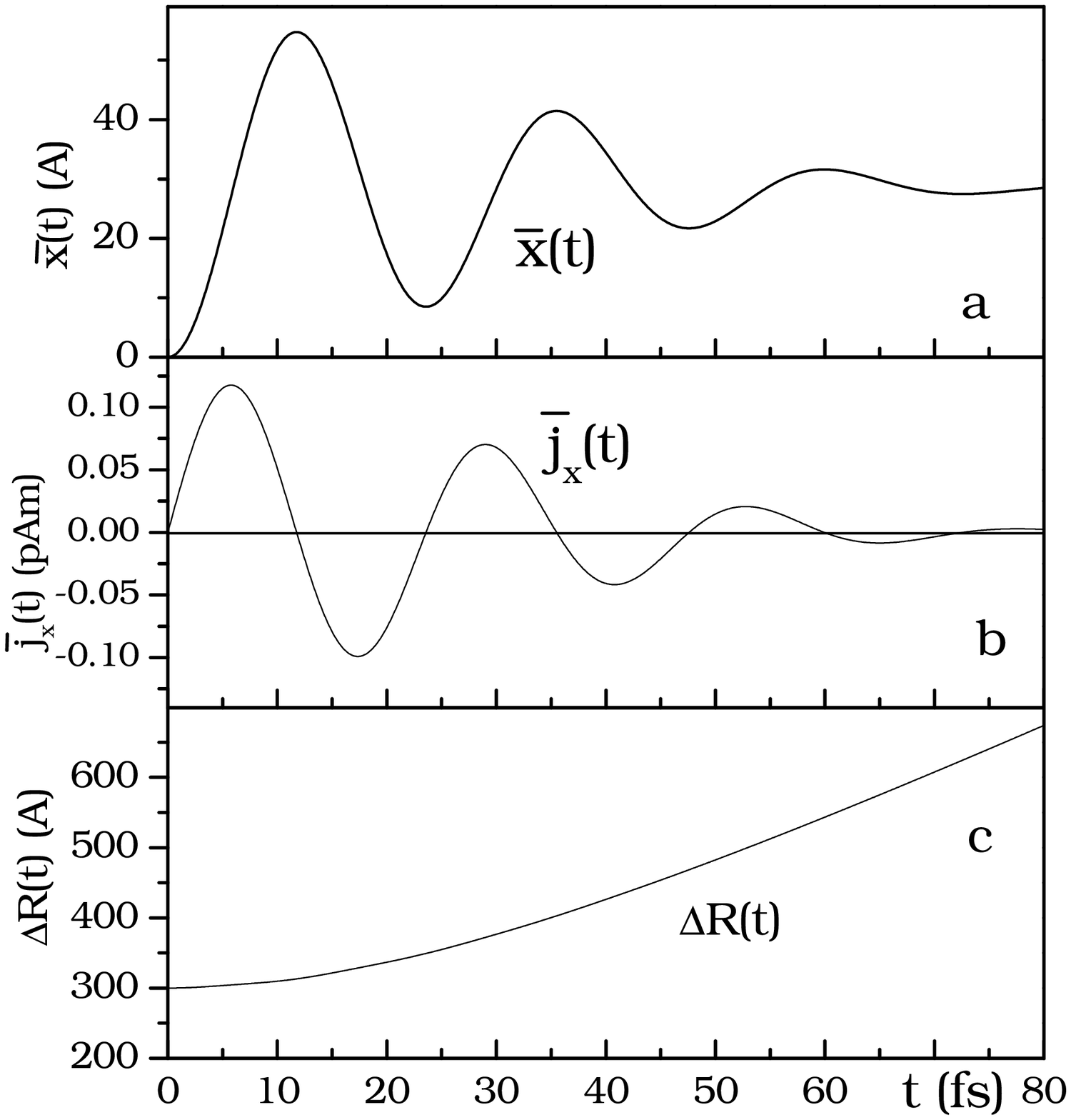}
\caption{\label{ZitterGrLFig1} Zitterbewegung of a charge carrier in bilayer graphene {\it versus}
time, calculated for a gaussian wave packet width $d=300$\AA\ and $k_{0y}=3.5\times 10^8$m$ ^{-1}$:
a) position, b) electric current,
c) dispersion $\Delta R(t)$. After the ZB disappears a constant shift remains. After~\cite{Rusin2007b}.}
\end{figure}

We first present the results for bilayer
graphene since they can be obtained in the analytical form, which allows one to see directly important
features of the trembling motion.
Two-dimensional Hamiltonian for bilayer graphene is well approximated by~\cite{McCann2006}
\begin{equation} \label{BG_H}
 \hat{H}_{B} = -\frac{1}{2m^*}\left(\begin{array}{cc}
     0 & (\hat{p}_x-i\hat{p}_y)^2 \\ (\hat{p}_x+i\hat{p}_y)^2 & 0 \end{array}\right),
\end{equation}
where $m^*=0.054m_0$. The energy spectrum is ${\cal E}_k=\pm E_k$,
where $E_k=\hbar^2k^2/2m^*$, i.e.
there is no energy gap between the conduction and valence bands. The position operator in the
Heisenberg picture is a $2\times 2$ matrix
$\hat{x}(t)=\exp(i\hat{H}_Bt/\hbar)\hat{x}\exp(-i\hat{H}_Bt/\hbar).$
One calculates
\begin{equation} \label{BG_x_11}
 x_{11}(t) = x(0) + \frac{k_y}{k^2}\left[ 1 -\cos\left(\frac{\hbar k^2t}{m^*}\right)\right],
\end{equation}
where $k^2=k_x^2+k_y^2$.
The third term represents the Zitterbewegung with the frequency
$\hbar\omega_Z=2\hbar^2k^2/2m^*$, corresponding to the energy difference
between the upper and lower energy branches for a given value of~$k$.
We want to calculate ZB of a charge carrier represented by a two-dimensional wave packet
centered at $\bm k_0=(0,k_{0y})$ and characterized by the width~$d$.
An average of $\hat{x}_{11}(t)$ is a
two-dimensional integral which can be calculated analytically
\begin{equation} \label{BG_x(t)}
\bar{x}_{11}(t) = \langle \psi(\bm {r})|\hat{x}(t)| \psi(\bm {r})\rangle
           = \bar{x}_c + \bar{x}_Z(t)
\end{equation}
where $\bar{x}_c=(1/k_{0y}) \left[1-\exp(-d^2k_{0y}^2)\right]$, and
\begin{eqnarray} \label{BG_xZ(t)}
\bar{x}_Z(t) = \frac{1}{k_{0y}} \left[ \exp\left(-\frac{\delta^4 d^2 k_{0y}^2}{d^4+\delta^4}\right)
               \cos\left(\frac{\delta^2d^4k_{0y}^2}{d^4+\delta^4} \right) \right. \nonumber \\
               \left. -\exp(-d^2k_{0y}^2) \right], \ \ \ \ \
\end{eqnarray}
in which $\delta=\sqrt{\hbar t/m^*}$ contains the time dependence.
In Figure~\ref{ZitterGrLFig1}a we show the ZB of the electron position $\bar{x}_{11}$
as given in~(\ref{BG_x(t)}) and~(\ref{BG_xZ(t)}).

We enumerate the main features of ZB
following from~(\ref{BG_x(t)}) and~(\ref{BG_xZ(t)}). First, in order to have
ZB in the direction~$x$ one needs an initial transverse momentum~$\hbar k_{0y}$.
Second, the ZB frequency depends only weakly on the packet width:
$\omega_Z=(\hbar k_{0y}^2/m^*)(d^4/(d^4+\delta^4))$, while its amplitude is strongly dependent
on the width~$d$. Third, the ZB has a transient character
since it is attenuated by the exponential term. For small~$t$ the amplitude of~$\bar{x}_Z(t)$ diminishes
as $\exp(-\Gamma_Z^2t^2)$ with
\begin{equation} \label{BG_GammaZ}
\Gamma_Z = \frac{\hbar k_{0y}}{m^*d}.
\end{equation}
Fourth, as~$t$ (or~$\delta$) increases,
the cosine term tends to unity and the first term in~(\ref{BG_xZ(t)}) cancels out with the second term,
which illustrates the Riemann-Lebesgue theorem (see~\cite{Lock1979}).
After the oscillations disappear, the charge carrier is displaced by the amount~$\bar{x}_c$, which is a
``remnant'' of ZB. Fifth, for very wide packets ($d\rightarrow\infty$) the exponent in~(\ref{BG_xZ(t)})
tends to unity, the oscillatory term is $\cos(\delta^2k_{0y}^2)$ and the last term vanishes.
In this limit one recovers undamped ZB oscillations.

Next, we consider other quantities related to ZB, beginning by the current.
The latter is given by the velocity multiplied by charge.
The velocity is simply $\bar{v}_x=\partial \bar{x}_Z/\partial t$,
where~$\bar{x}_Z$ is given by~(\ref{BG_xZ(t)}).
The calculated current is plotted in Figure~\ref{ZitterGrLFig1}b,
its oscillations are a direct manifestation of ZB.
The transient character of ZB is accompanied by a temporal spreading of the wave packet.
In fact, the question arises whether the attenuation of ZB is not simply {\it caused} by the
spreading of the packet.
The calculated packet width~$\Delta R$ is plotted versus time in Figure~\ref{ZitterGrLFig1}c. It is seen
that during the initial 80 femtoseconds the packet's width increases
only twice compared to its initial value,
while the ZB disappears almost completely. We conclude
that the spreading of the packet is {\it not} the main cause of the transient character of ZB.
Looking for physical reasons behind the transient character of ZB
we decompose the total wave function $\psi(\bm r,t)$ into the positive~($p$) and
negative~($n$) components $\psi^p(\bm r,t)$ and $\psi^n(\bm r,t)$.
We have
\begin{eqnarray} \label{BG_psit_S}
|\psi(t)\rangle &=& e^{-i\hat{H}t/\hbar} |\psi(0)\rangle \nonumber \\
 &=& e^{-iEt/\hbar}\langle p|\psi(0)\rangle |p\rangle +
      e^{iEt/\hbar}\langle n|\psi(0)\rangle |n\rangle, \ \
\end{eqnarray}
where $|p\rangle$ and $|n\rangle$ are the eigen-functions of the Hamiltonian~(\ref{BG_H})
in~$\bm k$ space corresponding to positive and negative energies, respectively.
Further
\begin{eqnarray}
 \langle \bm k|p\rangle &=& \frac{1}{\sqrt{2}}\left(\begin{array}{c} 1 \\
            k_+^2/k^2 \end{array}\right) \delta(\bm k-\bm k^{'}),\\
 \langle \bm k|n\rangle &=& \frac{1}{\sqrt{2}}\left(\begin{array}{c} 1 \\
           -k_+^2/k^2 \end{array}\right)\delta(\bm k-\bm k^{'}).
\end{eqnarray}
After some manipulations one obtains
\begin{eqnarray} \label{BG_psit_pos}
 \psi^p(\bm r,t) = \frac{1}{4\pi}\frac{d}{\sqrt{\pi}}
    \int d^2 \bm k e^{-\frac{1}{2}d^2(k_x^2+(k_y-k_{0y})^2)} e^{i\bm k\bm r} e^{-iEt/\hbar}
    \times \nonumber \\ \ \ \
    \left(\begin{array}{c} 1 \\ k_+^2/k^2 \end{array}\right). \ \ \ \ \ \
\end{eqnarray}
The function $\psi^n(\bm r,t)$ is given by the identical expression with the changed
signs in front of~$E$ and $k_+^2/k^2$ terms. There is
$\psi(\bm r,t)=\psi^p(\bm r,t)+\psi^n(\bm r,t)$ and $\langle \psi^n|\psi^p\rangle=0$.
Now, one can calculate the average values of $\bar{x}$ and $\bar{y}$ using the positive
and negative components in the above sense. We have
\begin{equation}
 \bar{x}(t) = \int (\psi^n + \psi^p)^{\dagger} x (\psi^n + \psi^p) d^2 \bm r,
\end{equation}
so that we deal with four integrals. A direct calculation gives
\begin{equation} \label{BG_x_mix}
 \int |\psi^p|^2x d^2 \bm r + \int |\psi^n|^2x d^2 \bm r = \bar{x}_c,
\end{equation}
\begin{equation}
 \int \psi^{n\dagger}x \psi^p d^2 \bm r + \int \psi^{p\dagger}x \psi^n d^2 \bm r = \bar{x}_Z(t),
\end{equation}
where $\bar{x}_c$ and $\bar{x}_Z(t)$ have been defined in~(\ref{BG_x(t)}).
Thus the integrals involving only the positive and only the negative components
give the constant shift due to ZB, while the mixed terms lead to the ZB oscillations.
All terms together reconstruct the result~(\ref{BG_x(t)}).
Next we calculate the average value~$\bar{y}$. There is no symmetry
between~$\bar{x}$ and~$\bar{y}$ because the wave packet is centered around
$k_x=0$ and $k_y=k_{0y}$. The average value~$\bar{y}$ is again given
by four integrals. However, now the mixed terms vanish,
while the integrals involving the positive and negative components give
\begin{eqnarray} \label{BG_psip}
 \int |\psi^p|^2y d^2 \bm r &=& \frac{\hbar k_{0y}}{2m^*}t, \\
 \label{BG_psin}
 \int |\psi^n|^2y d^2 \bm r &=& -\frac{\hbar k_{0y}}{2m^*}t.
\end{eqnarray}
This means that the ``positive'' and ``negative'' subpackets move in the opposite
directions with the velocity $v=\hbar k_{0y}t/2m^*$. The relative velocity
is $v^{rel}=\hbar k_{0y}t/m^*$. Each of these packets has the initial width~$d$
and it (slowly) spreads in time. After the time $\Gamma_Z^{-1}=d/v^{rel}$
the distance between the two packets equals~$d$, so the integrals~(\ref{BG_x_mix})
are small, resulting in the diminishing Zitterbewegung amplitude. This reasoning
gives the decay constant $\Gamma_Z=\hbar k_{0y}/m^*d$, which is exactly
what we determined above from the analytical results (see~(\ref{BG_GammaZ})).
Thus, {\it the transient character of the ZB oscillations is due to the
increasing spatial separation of the subpackets} corresponding to the positive and negative
energy states. This confirms our previous conclusion that it is not the packet's
slow spreading that is responsible for the attenuation.
The separation of subpackets with the resulting decay of ZB turns out to be
a general feature of this phenomenon.

\subsection{Monolayer graphene}

\begin{figure}
\includegraphics[width=8.5cm,height=8.5cm]{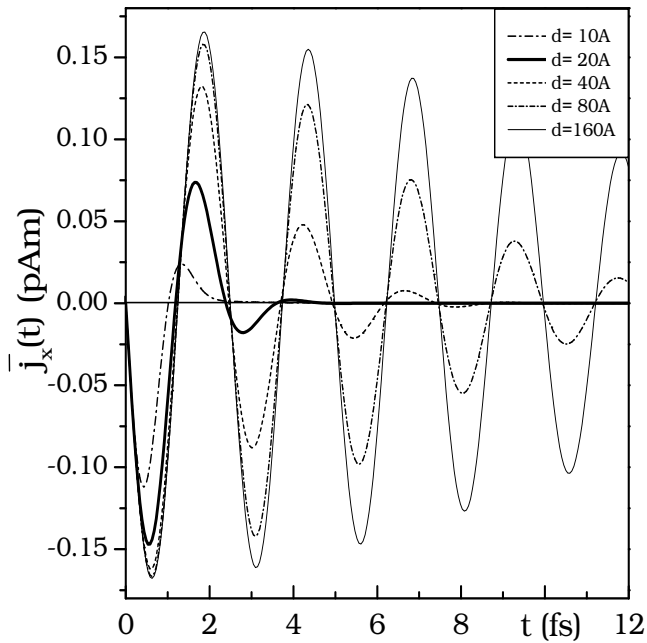}
\caption{Oscillatory electric current in the~$x$ direction caused by the ZB in monolayer graphene
{\it versus} time, calculated for a gaussian wave packet with $k_{0y}=1.2\times 10^9$m$^{-1}$ and various
packet widths~$d$. Transient character of ZB is clearly seen. After~\cite{Rusin2007b}.} \label{ZitterGrLFig2}
\end{figure}

Now we turn to monolayer graphene. The two-dimensional band Hamiltonian describing its band
structure is~\cite{Wallace1947,Slonczewski1958,Semenoff1984,Novoselov2004}
\begin{equation} \label{MG_H}
 \hat{H}_M = u\left(\begin{array}{cc}
     0 & \hat{p}_x-i\hat{p}_y \\ \hat{p}_x+i\hat{p}_y & 0 \end{array}\right),
\end{equation}
where $u\approx 1\times 10^8$cm/s. The resulting energy dispersion is linear in momentum:
${\cal E}=\pm u\hbar k$, where $k=\sqrt{k_x^2+k_y^2}$.
The quantum velocity in the Schrodinger picture is
$\hat{v}_i=\partial H_M/\partial \hat{p}_i$, it does not commute with the Hamiltonian~(\ref{MG_H}).
In the Heisenberg picture we have
$\hat{\bm v}(t)=\exp(i\hat{H}_Mt/\hbar){\bm \hat{v}}\exp(-i\hat{H}_Mt/\hbar)$.
Using~(\ref{MG_H}) one calculates
\begin{equation} \label{MG_v11}
 v_x^{(11)} = u \frac{k_y}{k}\sin(2ukt).
\end{equation}
The above equation describes the trembling motion with the frequency $\omega_Z=2uk$,
determined by the energy difference between the upper and lower energy branches for a given
value of~$k$. As before, ZB in the direction~$x$ occurs only if there is a non-vanishing
momentum $\hbar k_y$. One calculates an average velocity (or current) taken over a two-dimensional wave
packet with nonzero initial momentum~$k_{0x}$. The results for the current
$\bar{j}_x=e\bar{v}_x$ are plotted in Figure~\ref{ZitterGrLFig2} for different realistic packet widths~$d$.
It is seen that the ZB frequency does not depend
on~$d$ and is nearly equal to~$\omega_Z$ given above for the plane wave.
On the other hand, the amplitude of ZB does depend on~$d$ and we deal with decay times
of the order of femtoseconds. For small~$d$ there are almost no oscillations, for
very large~$d$ the ZB oscillations are undamped. These conclusions agree with our analytical
results for bilayer graphene. The behavior of ZB depends quite critically on the
values of~$k_{0y}$ and~$d$, which is reminiscent of the damped harmonic oscillator.
In the limit $d\rightarrow \infty$ the above results for the electric current resemble those of
Katsnelson~\cite{Katsnelson2006} for ZB in graphene obtained with the use of plane wave representation.

Maksimova {\it et al.}~\cite{Maksimova2008} investigated dynamics of wave packets in monolayer graphene
for different pseudo-spin polarizations with the resulting ZB. For specific packet components and their
relative phases a ``longitudinal ZB'' can take place, but its intensity is weak.

Martinez {\it et al.}~\cite{Martinez2010} considered a creation of electron-hole pairs by a constant
electric field in the plane of a monolayer graphene sheet. They showed that, as the pairs
undergo the ZB in opposite directions, a Hall-like separation of the charge occurs
giving a measurable dipole moment.
We note that it is not the time-dependent motion but the ZB shift at large times which is responsible
for the charge separation, see~(\ref{BG_x_11}) and Figure~\ref{ZitterGrLFig1}.

Englman and Vertesi~\cite{Englman2008} calculated a ZB-related electron current in monolayer graphene
in the adiabatic approximation and related it to the Berry phase.

\subsection{Carbon nanotubes}
\begin{figure}
\includegraphics[width=8.5cm,height=8.5cm]{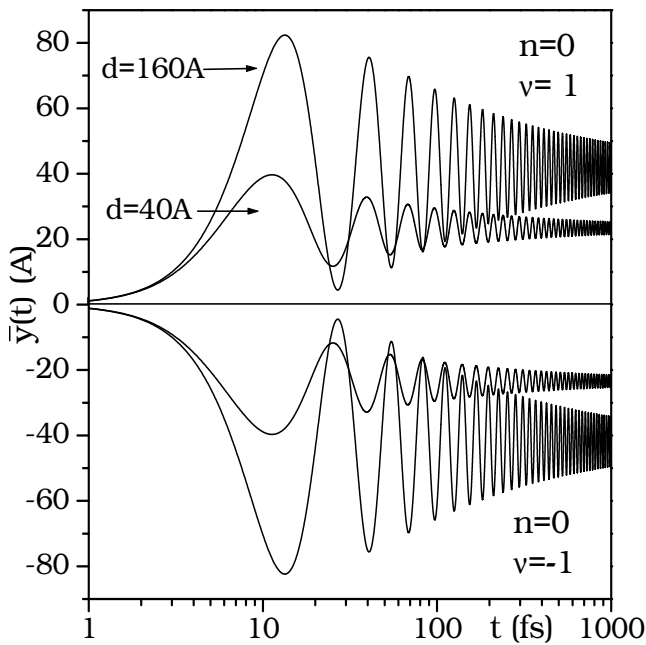}
\caption{Zitterbewegung of two charge carriers in the ground subband of a single carbon
nanotube of $L=200$ \AA\ {\it versus} time (logarithmic scale),
calculated for gaussian wave packets of two different
widths~$d$ and~$k_{0y}=0$. After the ZB disappears a constant shift remains.
The two carriers are described by different quantum numbers~$\nu$.
At higher times the amplitude of ZB oscillations decays as~$t^{-1/2}$. After~\cite{Rusin2007b}.}
\label{ZitterGrLFig3}
\end{figure}

Next, we consider monolayer graphene sheets rolled into single
semiconducting carbon nanotubes~(CNT)~\cite{Zawadzki2006,Rusin2007b}.
The band Hamiltonian in the vicinity of~$K$ point is~\cite{Ajiki93}
\begin{equation} \label{MG_CNT}
 \hat{H}_{CNT} = u\left(\begin{array}{cc}
     0 & \hbar k_{n\nu}-i\hat{p}_y \\ \hbar k_{n\nu}+i\hat{p}_y & 0 \end{array}\right).
\end{equation}
This Hamiltonian is similar to~(\ref{MG_H}) except that, because of
the periodic boundary conditions, the momentum~$p_x$ is quantized
and takes discrete values $\hbar k_x=\hbar k_{n\nu}$, where $k_{n\nu}=(2\pi/L)(n-\nu/3)$, $n=0,\pm 1,\ldots$,
$\nu=\pm 1$, and~$L$ is the length of circumference of CNT.
As a result, the free electron motion can occur only in the direction
$y$, parallel to the tube axis. The geometry of CNT has
important consequences.
There exists an energy gap $E_g=2u\hbar|k_{n\nu}|$ and the effective mass at
the band edge $m_0^*=\hbar|k_{n\nu}|/u$. For $\nu=\pm 1$ there {\it always}
exists a non-vanishing value of the quantized momentum $\hbar k_{n\nu}$.
Finally, for each value of~$k_{n\nu}$ there exists $k_{-n,-\nu}=-k_{n\nu}$ resulting in the
same subband energy ${\cal E}=\pm E$, where
\begin{equation} \label{CNT_E}
 E=\hbar u\sqrt{k_{n\nu}^2+k_y^2}.
\end{equation}
The time dependent velocity $\hat{v}_y(t)$ and the displacement $\hat{y}(t)$ can be
calculated for the plane electron wave in the usual way and they exhibit the ZB
oscillations (see~\cite{Zawadzki2006}). For small momenta~$k_y$ the ZB frequency
is $\hbar\omega_Z=E_g$ and the ZB length is $\lambda_Z\approx 1/|k_{n\nu}|$.
We are again interested in the displacement $\bar{y}(t)$ of a charge carrier
represented by a one-dimensional wave packet analogous to that described in~(\ref{defPacket})
The average displacement is $\bar{y}(t)=\bar{y}_Z(t)-\bar{y}_{sh}$, where
\begin{equation} \label{CNT_y_osc}
\bar{y}_Z(t) = \frac{\hbar^2du^2k_{n\nu}}{2\sqrt{\pi}} \int_{-\infty}^{\infty}
    \frac{dk_y} {E^2}\cos\left(\frac{2Et}{\hbar}\right)e^{-d^2k_y^2}
\end{equation}
and $\bar{y}_{sh}=1/2\sqrt{\pi}d\ {\rm sgn}(b)[1-\Phi(|b|)]\exp(b^2)$, where $b=k_{n\nu}d$ and
$\Phi(x)$ is the error function. The ZB oscillations of $\bar{y}(t)$ are plotted in Figure~\ref{ZitterGrLFig3}.
It is seen that, after the transient ZB oscillations disappear, there
remains a shift $\bar{y}_{sh}$. Thus the ZB separates spatially the charge carriers that are
degenerate in energy but characterized by $n,\nu$ and $-n,-\nu$ quantum numbers.
The current is proportional to $\bar{v}_y=\partial \bar{y}/\partial t$, so that the currents
related to $\nu=1$ and $\nu=-1$ cancel each other. To have a non-vanishing current one needs to break the
above symmetry, which can be achieved by applying an external magnetic field
parallel to the tube axis.

It can be seen from Figure~\ref{ZitterGrLFig3} that the decay time of ZB in CNT is much larger than that in bilayer
and monolayer graphene. The oscillations decrease proportionally to~$t^{-1/2}$.
The reason is that we consider the situation with $k_{0y}=0$, so that the ZB
oscillations occur due to ``built in'' momentum $k_x=k_{n\nu}$, arising from the tube's topology. In other words,
the long decay time is due to the one-dimensionality of the system. If the circumference of a CNT is
increased, the energy gap (and, correspondingly, the ZB frequency) decreases, the amplitude of ZB
is larger, but the decay time remains almost unchanged.

One can show that we again deal here with two sub-packets
which, however, for $k_{0y}=0$ do not run away from each other. Thus, the slow damping of ZB
is due only to the slow broadening of the sub-packets. We emphasize the slow decay, as
illustrated in Figure~\ref{ZitterGrLFig3}, because
it is confirmed experimentally, see Section~VIII. We add that for $k_{0y} \neq 0$ the sub-packets
run away from each other and the decay time is much faster.

\section{ZB in a magnetic field}

The trembling motion of charge carriers in solids has been described
above for no external potentials. Now we consider the trembling motion of electrons in
the presence of an external magnetic field~\cite{Rusin2008}.
The magnetic field is known to cause no
interband electron transitions, so the essential features of ZB are expected not to be destroyed.
On the other hand, introduction of an external field provides an important
parameter affecting the ZB behavior. This case is special because the electron spectrum is fully
quantized.
We consider a graphene monolayer in an external magnetic field parallel to the~$z$ axis.
The Hamiltonian for electrons and holes at the~$K_1$ point is~\cite{Wallace1947,Slonczewski1958}
\begin{equation} \label{H_pi0}
\hat{H} = u\left(\begin{array}{cc}
     0 & \hat{\pi}_x-i\hat{\pi}_y \\ \hat{\pi}_x+i\hat{\pi}_y & 0 \end{array}\right),
\end{equation}
where $u\approx 1\times $10$^8$ cm/s is the characteristic velocity,
$\hat{\bm \pi}=\hat{\bm p}-q \hat{\bm A}$ is the generalized momentum, in which
$\hat{\bm A}$ is the vector potential and~$q$ is the electron charge.
Using the Landau gauge, we take $\hat{\bm A}=(-By,0,0)$,
and for an electron~$q=-e$ with~$e>0$. We take the wave
function in the form $\Psi(x,y)=e^{ik_xx}\Phi(y)$.
Introducing the magnetic radius $L=\sqrt{\hbar/eB}$, the variable $\xi=y/L-k_xL$, and
defining the standard raising and lowering operators for the harmonic oscillator
$\hat{a}=(\xi+\partial/\partial \xi)/\sqrt{2}$ and $\hat{a}^{\dagger}=(\xi-\partial/\partial \xi)/\sqrt{2}$,
the Hamiltonian becomes
\begin{equation} \label{H aap}
 \hat{H} = -\hbar\Omega\left(\begin{array}{cc} 0 & \hat{a} \\ \hat{a}^{\dagger}& 0 \\ \end{array}\right),
\end{equation}
where the frequency is $\Omega=\sqrt{2}u/L$.
Next one determines the eigenstates and eigenenergies of the Hamiltonian~$\hat{H}$.
The energy is $E_{ns}=s\hbar\Omega\sqrt{n}$.
Here $n=0,1,\ldots$, and $s=\pm 1$ for the conduction and valence bands, respectively.
The above energies were confirmed experimentally. The complete wave function is
\begin{equation} \label{H_nskx}
|{\rm n}\rangle \equiv |nk_xs\rangle = \frac{e^{ik_xx}}{\sqrt{4\pi}}
               \left(\begin{array}{c} -s|n-1\rangle \\|n\rangle \end{array}\right)
\end{equation}
where $|n\rangle$ are the harmonic oscillator functions.

\begin{figure}
\includegraphics[width=8.5cm,height=8.5cm]{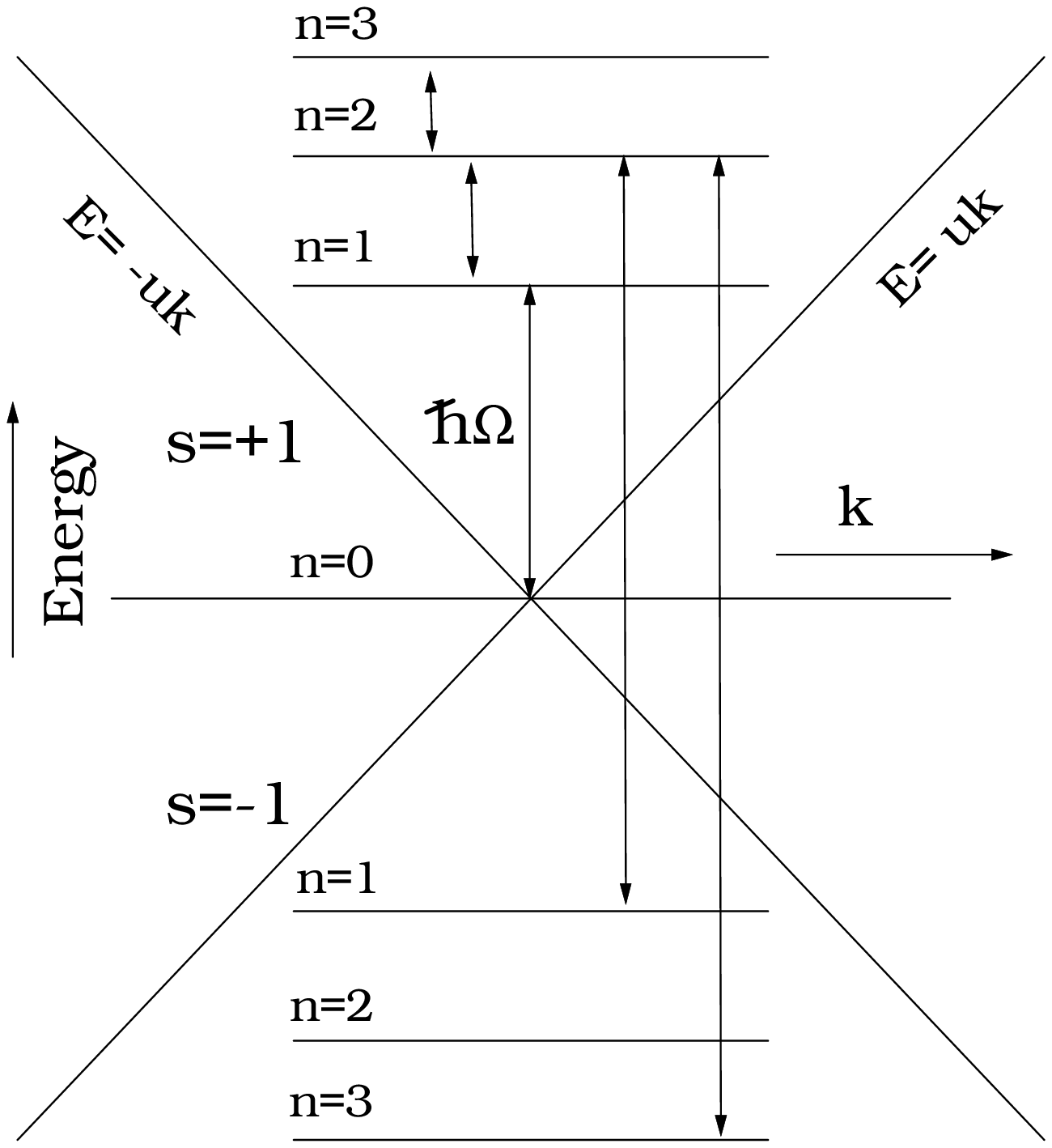}
\caption{ \label{GraphHFig1} The energy dispersion~$E(k)$ and the Landau levels for monolayer graphene in a
magnetic field (schematically). Intraband (cyclotron) and interband (ZB) energies for $n'=n\pm 1$
are indicated. The basic energy is $\hbar\Omega=\sqrt{2}\hbar u/L$. After~\cite{Rusin2008}.}
\end{figure}

We want to calculate the velocity of charge carriers described by a wave packet. We first calculate
matrix elements $\langle f|{\rm n}\rangle$ between an arbitrary
two-component function $f=(f^u,f^l)$ and eigenstates~(\ref{H_nskx}). A straightforward
manipulation gives $\langle f|{\rm n}\rangle=-sF^u_{n-1}+F^l_{n}$, where
\begin{equation} \label{V_Matrix_Fj}
  F^j_n(k_x) = \frac{1}{\sqrt{2L}C_n} \int g^j(k_x,y)e^{-\frac{1}{2}\xi^2}{\rm H}_{n}(\xi)dy,
\end{equation}
in which
\begin{equation} \label{V_Matrix_gj}
 g^j(k_x,y) = \frac{1}{\sqrt{2\pi}} \int f^j(x,y)e^{ik_xx} dx.
\end{equation}
The superscript $j=u,l$ stands for the upper and lower components of the function~$f$.
The Hamilton equations give the velocity components:
$\hat{v}_i(0)=\partial \hat{H}/\partial \hat{p}_i$, with $i=x,y$. We want to
calculate averages of the time-dependent velocity operators $\hat{v}_i(t)$ in the
Heisenberg picture taken on the function~$f$. The averages are
\begin{equation} \label{V_vi(t)}
 \bar{v}_i(t) = \sum_{\rm n,n'} e^{iE_{\rm n'}t/\hbar}\langle f|{\rm n'}\rangle \langle {\rm n'}|v_i(0)|\rm n \rangle
                \langle {\rm n}|f\rangle e^{-iE_{\rm n}t/\hbar},
\end{equation}
where the energies and eigenstates are given in~(\ref{H_nskx}). The summation in~(\ref{V_vi(t)}) goes over
all the quantum numbers: $n,n',s,s',k_x,k_x'$.
The only non-vanishing matrix elements of the velocity components are for the states
states $n'=n\pm 1$. One finally obtains after some manipulation
\begin{widetext} \begin{eqnarray} \label{V_vy(t)}
\bar{v}_y(t) &=&  u\sum_{n=0}^{\infty} V_n^+\sin(\omega_n^ct) +  u\sum_{n=0}^{\infty}V_n^-\sin(\omega_n^Zt)
                +iu\sum_{n=0}^{\infty} A_n^+\cos(\omega_n^ct) + iu\sum_{n=0}^{\infty}A_n^-\cos(\omega_n^Zt),\\
                \label{V_vx(t)}
\bar{v}_x(t) &=& u \sum_{n=0}^{\infty} B_n^+\cos(\omega_n^ct) +  u\sum_{n=0}^{\infty}B_n^-\cos(\omega_n^Zt)
                +iu\sum_{n=0}^{\infty} T_n^+\sin(\omega_n^ct) + iu\sum_{n=0}^{\infty}T_n^-\sin(\omega_n^Zt),
\end{eqnarray} \end{widetext}
where $V_n^{\pm}$, $T_n^{\pm}$, $A_n^{\pm}$ and $B_n^{\pm}$ are given by combinations of
$U^{\alpha,\beta}_{m,n}$ integrals
\begin{equation} \label{V_def_U}
U^{\alpha,\beta}_{m,n} = \int F_{m}^{\alpha *}(k_x) F_{n}^{\beta}(k_x) dk_x.
\end{equation}
The superscripts~$\alpha$ and~$\beta$ refer to the upper and lower components, see~\cite{Rusin2008}.
The time dependent sine and cosine functions come from
the exponential terms in~(\ref{V_vi(t)}).
The frequencies in~(\ref{V_vy(t)}) and~(\ref{V_vx(t)})
are $\omega_n^c=\Omega(\sqrt{n+1}-\sqrt{n})$,
$\omega_n^Z=\Omega(\sqrt{n+1}+\sqrt{n})$, where~$\Omega$ is given in~(\ref{H aap}).
The frequencies~$\omega_n^c$ correspond to the
intraband energies while frequencies~$\omega_n^Z$ correspond to the interband energies, see
Figure~\ref{GraphHFig1}.
The interband frequencies are characteristic of the Zitterbewegung.
The intraband (cyclotron) energies are due to the band quantization by the magnetic field and they
do not appear in field-free situations.

Final calculations were carried out for a two-dimensional Gaussian wave packet centered around the
wave vector ${\bf k_0}=(k_{0x},0)$ and having two non-vanishing components.
In this case one can obtain analytical expressions for $U^{\alpha,\beta}_{m,n}$.
The main frequency of oscillations is $\omega_0=\Omega$, which can be
interpreted either as $\omega_0^c=\Omega(\sqrt{n+1}-\sqrt{n})$ or
$\omega_0^Z=\Omega(\sqrt{n+1}+\sqrt{n})$
for~$n=0$. Frequency $\omega_0^c$ belongs to the intraband (cyclotron) set, while~$\omega_0^Z$
belongs to the interband set (see Figure~\ref{GraphHFig1}).
The striking feature is, that ZB is
manifested by several frequencies simultaneously. This is a consequence of the fact that
in graphene the energy distances between the Landau levels diminish with~$n$,
which results in different values of frequencies~$\omega_n^c$ and~$\omega_n^Z$ for different~$n$.
It follows that it is the presence of an external quantizing magnetic field that
introduces various frequencies into ZB. It turns out that,
after the ZB oscillations seemingly die out, they actually reappear at higher times.
Thus, {\it for all $k_{0x}$ values (including $k_{0x}=0$), the ZB oscillations have a permanent character},
that is they do not disappear in time. This feature is due to the discrete character of the electron spectrum
caused by a magnetic field. The above property is in sharp contrast to the no-field cases
considered above, in which the spectrum is not quantized and the ZB of a wave packet
has a transient character.
In mathematical terms, due to the discrete character of the spectrum,
averages of operator quantities taken over a wave packet are sums and not integrals.
The sums do not obey the Riemann-Lebesgues theorem for
integrals which guaranteed the damping of ZB in time for a continuous spectrum (see~\cite{Lock1979}).

\begin{figure}
\includegraphics[width=8.5cm,height=8.5cm]{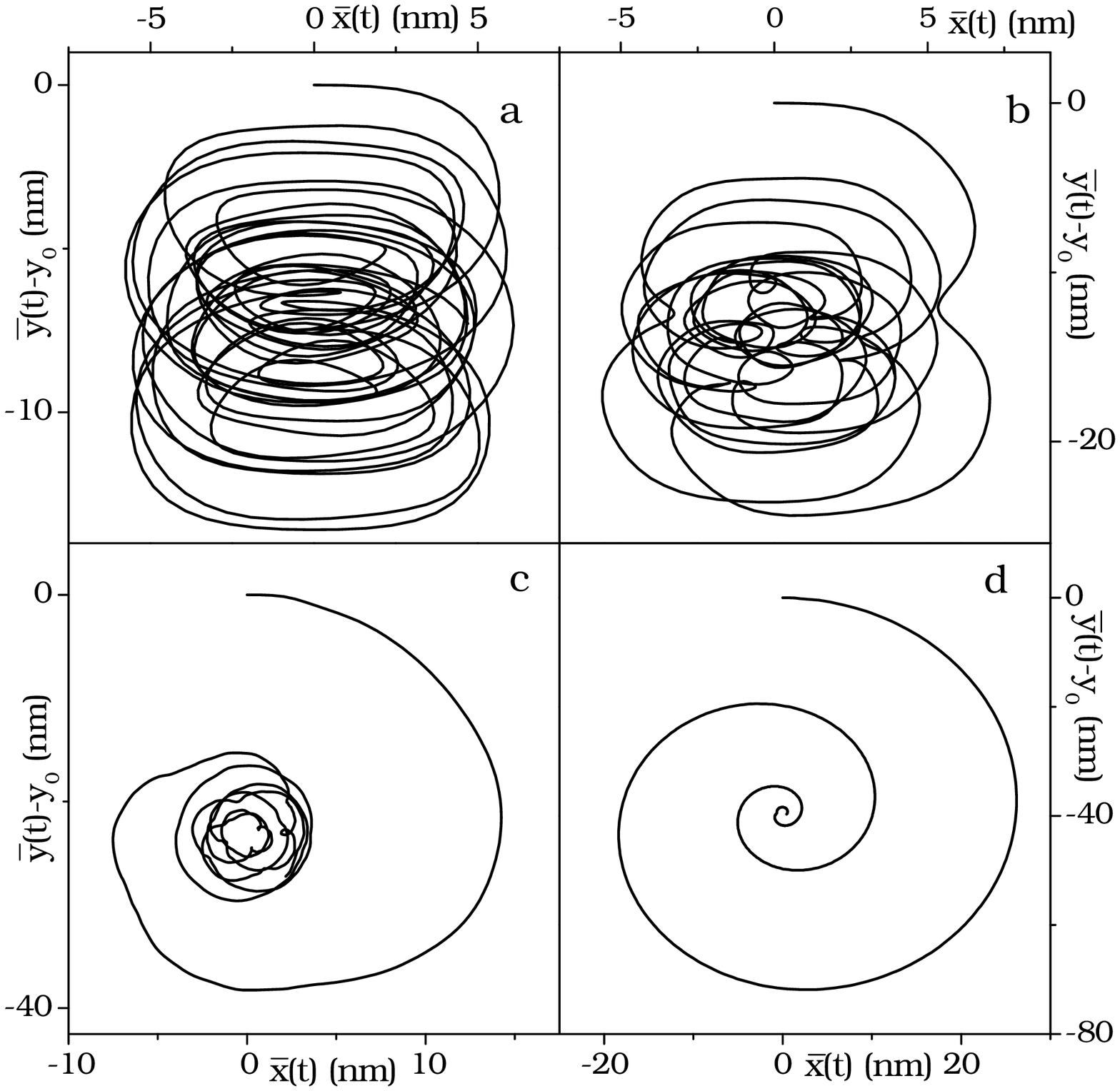}
\caption{ \label{GraphHFig5} Zitterbewegung trajectories of electron at the~$K_1$ point of the Brillouin
zone in monolayer graphene at B=20T during the first picosecond for various values of~$k_{0x}$.
After~\cite{Rusin2008}.}
\end{figure}

Finally, one calculates the displacements $\bar{x}(t)$ and $\bar{y}(t)$ of the wave packet. To this end we
integrate~(\ref{V_vy(t)}) and~(\ref{V_vx(t)}) with respect to time using the initial conditions
$x_0=\bar{x}(0)=0$ and $y_0=\bar{y}(0)=k_xL^2$. The results are plotted in Figure~\ref{GraphHFig5} in the
form of~ $x-y$ trajectories for different initial wave vectors~$k_{0x}$.
The direction of movement is clockwise and
the trajectories span early times (1ps) after the creation of a wave packet.

All in all, the presence of a quantizing magnetic field has the following important effects on the
trembling motion. (1) For $B\neq 0$ the ZB oscillation are permanent, while for~$B=0$ they are transient.
The reason is that for $B\neq 0$ the electron spectrum is discrete.
(2) For $B\neq 0$ many ZB frequencies appear, whereas for~$B=0$ only one ZB frequency exists.
(3) For $B\neq 0$ both interband and intraband (cyclotron) frequencies appear in ZB; for $B=0$ there
are no intraband frequencies. (4) Magnetic field intensity changes not only the ZB frequencies
but the entire character of ZB spectrum.

\begin{figure}
\includegraphics[width=8.5cm,height=8.5cm]{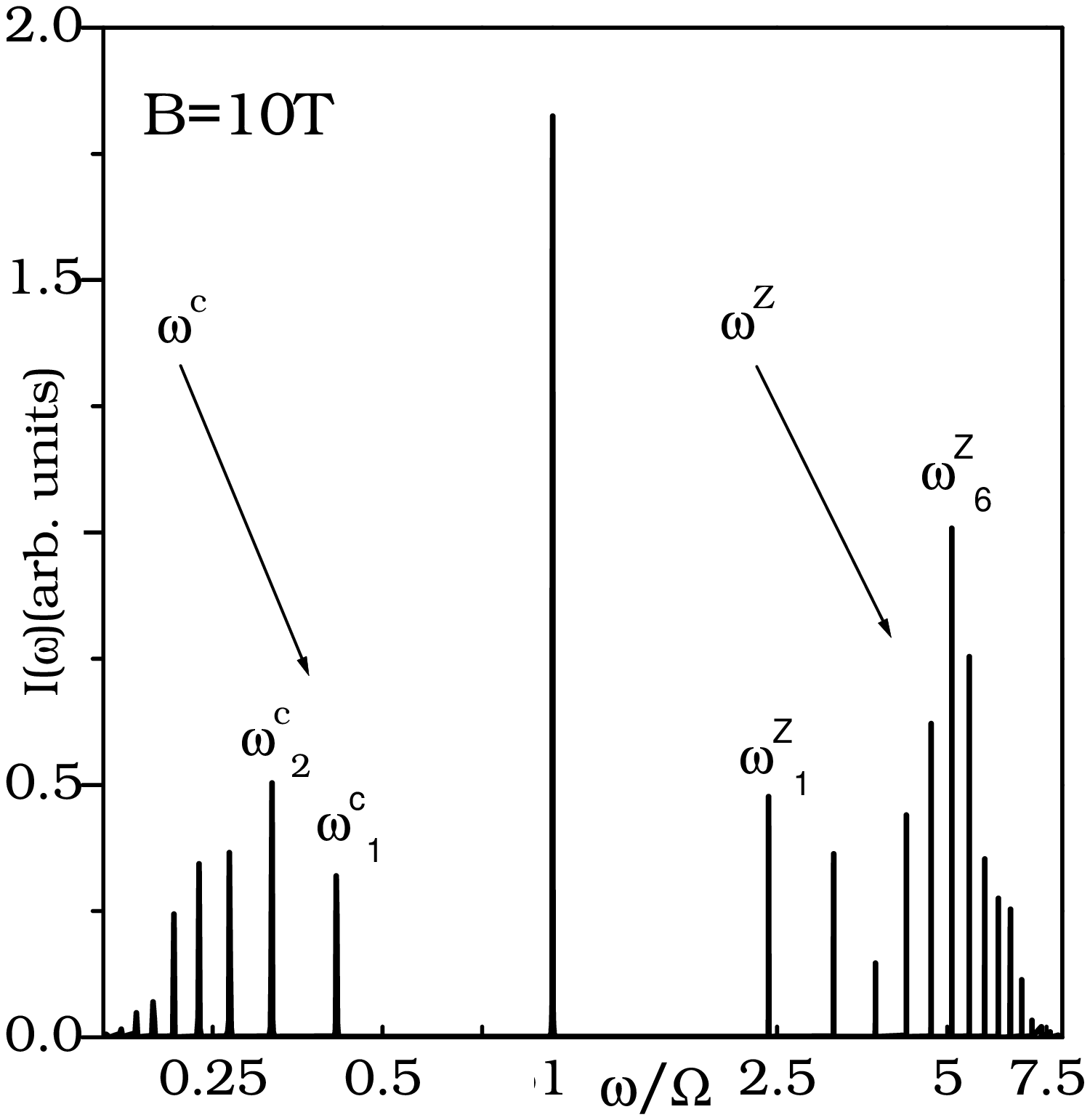}
\caption{ \label{GraphHFig6} Intensity spectrum versus frequency during the first 20 ps of motion of an
electron described by a Gaussian wave packet having
$k_{0x}=0.035$\AA$^{-1}$ in monolayer graphene. After~\cite{Rusin2008}.}
\end{figure}

The Zitterbewegung should be accompanied by electromagnetic dipole radiation emitted by the trembling
electrons. The oscillations $\bar{\bm r}(t)$ are related to the dipole moment $-e\bar{\bm r}(t)$,
which couples to the electromagnetic radiation. One can calculate the emitted electric field
from the electron acceleration $\bar{\ddot{\bm r}}(t)$ and takes its Fourier transform to determine
the emitted frequencies. In Figure~\ref{GraphHFig6} we plot the calculated intensities of various emitted lines.
The strong peak corresponds to oscillations with the basic frequency $\omega=\Omega$. The peaks on the
high-frequency side correspond to the interband excitations and are characteristic of ZB. The peaks
on the lower frequency side correspond to the intraband (cyclotron) excitations.
In absence of ZB the emission spectrum would contain only the intraband (cyclotron) frequencies.
Thus the interband frequencies $\omega_n^Z$ shown in Figure~\ref{GraphHFig6} are a direct signature of the trembling
motion. It can be seen that the $\omega_z^Z$ peaks are not much weaker than the central peak
at $\omega=\Omega$, which means that there exists a reasonable chance to observe them.
Generally speaking, the excitation of the system is due to the nonzero momentum $\hbar k_{0x}$ given
to the electron. It can be provided by accelerating the electron in the band or by exciting the
electron with a nonzero momentum by light from the valence band to the conduction band. The electron
can emit light because the Gaussian wave packet is not an eigenstate of the system described by the
Hamiltonian~(\ref{H_pi0}). The energy of the emitted light is provided by the initial kinetic energy related
to the momentum $\hbar k_{0x}$. Once this energy is completely used, the emission will cease. Radiation
emitted by the trembling electrons in monolayer graphene excited by femtosecond laser pulses is described
in Reference~\cite{Rusin2009}. This problem is not trivial since it is difficult to prepare an electron
in a solid in the form of a Gaussian wave packet. On the other hand, a formation of a light wave packet
is mastered by present technics. It was shown that, when the Landau levels are broadened by scattering or
defects, the light emission is changed from sustained to decaying in time.

Schlieman~\cite{Schliemann2008} described time dependence of the cyclotron motion in monolayer graphene
in the presence of a magnetic field using the semiclassical approximation for high carrier energies.
He showed that the cyclotron motion is perturbed by interband ZB contributions of higher frequencies.

Krueckl and Kramer~\cite{Krueckl2009} described time propagation of an initially
concentrated wave packet in monolayer graphene in a perpendicular magnetic field. A collapse-revival
pattern of ZB was investigated and an effect of impurities (disorder) on the packet dynamics
was analyzed. It turned out that ZB ``survives'' the perturbation by impurities.

Romera and de los Santos~\cite{Romera2009} studied monolayer graphene in a magnetic field
concentrating on collapse-revival pattern of ZB oscillations.

Wang {\it et al.}~\cite{Wang2009a} carried out a study similar to the one described above, but
for bilayer graphene in a magnetic field. This system is somewhat different from
monolayer graphene since the Landau levels are nearly uniformly spaced due to quadratic
dependence of positive and negative energies on momentum, see~(\ref{BG_H}). Also, the laser pulse
was assumed to contain only one frequency~$\omega_L$. The authors estimated that in high
quality bilayer graphene samples the stimulated ZB electric field can be of the order of
volts per meter and the corresponding coherence times of tens of femtoseconds.

Zulicke {\it et al.}~\cite{Zulicke2007} investigated the influence of ZB on the cyclotron motion considering
the so called Landau-Rashba Hamiltonian which, in addition to the 2D motion in a magnetic field,
contains also the Bychkov-Rashba spin-orbit term due to the structure inversion asymmetry. The latter
is a source of ZB, see~\cite{Schliemann2005}.

Demikhovskii {\it et al.}~\cite{Demikhovskii2008} studied 2D electron dynamics
in the presence of Bychkov-Rashba spin splitting. It was shown that in this case one deals with two
spin sub-packets propagating with unequal group velocities.
As the sub-packets go apart, their weakening interference is responsible for
a transient character of ZB in time. It was also demonstrated that in the presence of an external magnetic
field the spin sub-packets rotate with different cyclotron frequencies.

\section{Nature of ZB in Solids}
In spite of the great interest in the phenomenon of ZB its physical origin remained mysterious.
As mentioned above, it was recognized that the ZB in a vacuum is due to an interference of states
corresponding to positive and negative electron energies.
Since the ZB in solids was treated by the two-band Hamiltonian similar to the Dirac equation,
its interpretation was also similar. This did not explain its origin, it only provided a way to describe it.
For this reason we consider the fundamentals of electron propagation in a periodic potential
trying to elucidate
the nature of electron Zitterbewegung in solids. The physical origin of ZB is essential because
it resolves the question of its observability.
The second purpose is to decide whether the two-band {\bf k.p} model of the band structure,
used to describe the ZB in solids, is adequate.

One should keep in mind that we described above {\it various kinds} of ZB.
Every time one deals with two interacting energy bands, an interference of the lower and upper states
results in electron oscillations. In particular, one deals with ZB related to the
Bychkov-Rashba-type spin subbands~\cite{Schliemann2005} or to the Luttinger-type light and heavy hole
subbands~\cite{Winkler2007,Jiang2005}. However, the problem of our
interest here is the simplest electron propagation in a periodic potential. The trembling motion
of this type was first treated in~\cite{Zawadzki2005KP,Rusin2007b}.
It is often stated that an electron moving in a periodic potential behaves like a
free particle characterized by an effective mass~$m^*$. The above picture suggests that, if there
are no external forces, the electron moves in a crystal with a constant velocity. This, however, is
clearly untrue because the electron velocity operator $\hat{v}_i=\hat{p}_i/m_0$ does
not commute with the Hamiltonian $\hat{H}=\hat{\bf p}^2/2m_0+V({\bf r})$, so that~$\hat{v}_i$ is not a
constant of the motion. In reality, as the electron moves in a periodic potential, it
accelerates or slows down keeping its total energy constant. This situation is analogous to that
of a roller-coaster: as it goes down losing its potential energy, its velocity
(i.e. its kinetic energy) increases, and when it goes up its velocity decreases.

We first consider the trembling frequency~$\omega_Z$\cite{Zawadzki2010}. The latter is easy to determine
if we assume, in the first approximation, that the electron moves with a constant average velocity
$\bar{v}$ and the period of the potential is~$a$, so $\omega_Z=2\pi \bar{v}/a$.
Putting typical values for GaAs: $a=5.66$\AA, $\bar{v}=2.3 \times 10^7$cm/s,
one obtains $\hbar \omega_Z=1.68$eV, i.e. the interband frequency since
the energy gap is $E_g \simeq 1.5$eV. The interband frequency is in fact typical for the ZB in solids.

Next, we describe the velocity oscillations classically assuming for simplicity a one-dimensional
periodic potential of the form $V(z)=V_0\sin(2\pi z/a)$. The first integral of the motion expressing
the total energy is: $E=m_0v_z^2/2 + V(z)$. Thus the velocity is
\begin{equation} \label{vz}
\frac{dz}{dt} = \sqrt{\frac{2E}{m_0}}\left[1-\frac{V(z)}{E} \right]^{1/2}.
\end{equation}
One can now separate the variables and integrate each side in the standard way.
In the classical approach $V_0$ must be smaller than $E$. In general, the integration
of Eq.~(\ref{vz}) leads to elliptical integrals. However, trying
to obtain an analytical result we assume $V_0(z) \simeq E/2$, expand the square root
retaining the first two terms and solve the remaining equation by iteration taking in the first step
a constant velocity $v_{z0}=(2E/m_0)^{1/2}$.
This gives $z=v_{z0}t$ and
\begin{equation}
v_z(t) \simeq v_{z0}- \frac{v_{z0}V_0}{2E}\sin\left(\frac{2\pi v_{z0}t}{a}\right).
\end{equation}
Thus, as a result of the motion in a periodic potential, the electron velocity oscillates with the
expected frequency $\omega_Z=2\pi v_{z0}/a$ around the average value $v_{z0}$. Integrating
with respect to time we get an amplitude of ZB: $\Delta z = V_0a/(4\pi E)$.
Taking again $V_0\simeq E/2$, and estimating the lattice constant to be $a\simeq \hbar p_{cv}/(m_0E_g)$
(see Luttinger and Kohn~\cite{Luttinger1955}), we have finally $\Delta z \simeq \hbar p_{cv}/(8\pi m_0E_g)$,
where $p_{cv}$ is the interband matrix element of momentum. This should be compared with an
estimation obtained previously from the two-band {\bf k.p}
model~\cite{Zawadzki2005KP}: $\Delta z \simeq \lambda_Z = \hbar/m^*u=
\hbar(2/m^*E_g)^{1/2} \simeq 2\hbar p_{cv}/m_0E_g$. Thus the classical and
quantum results depend the same way on the fundamental parameters, although the classical
approach makes no use of the energy band structure. We conclude that the Zitterbewegung in solids
is simply due to the electron velocity oscillations assuring the energy conservation during
motion in a periodic potential.

\begin{figure}
\includegraphics[width=8cm,height=8cm]{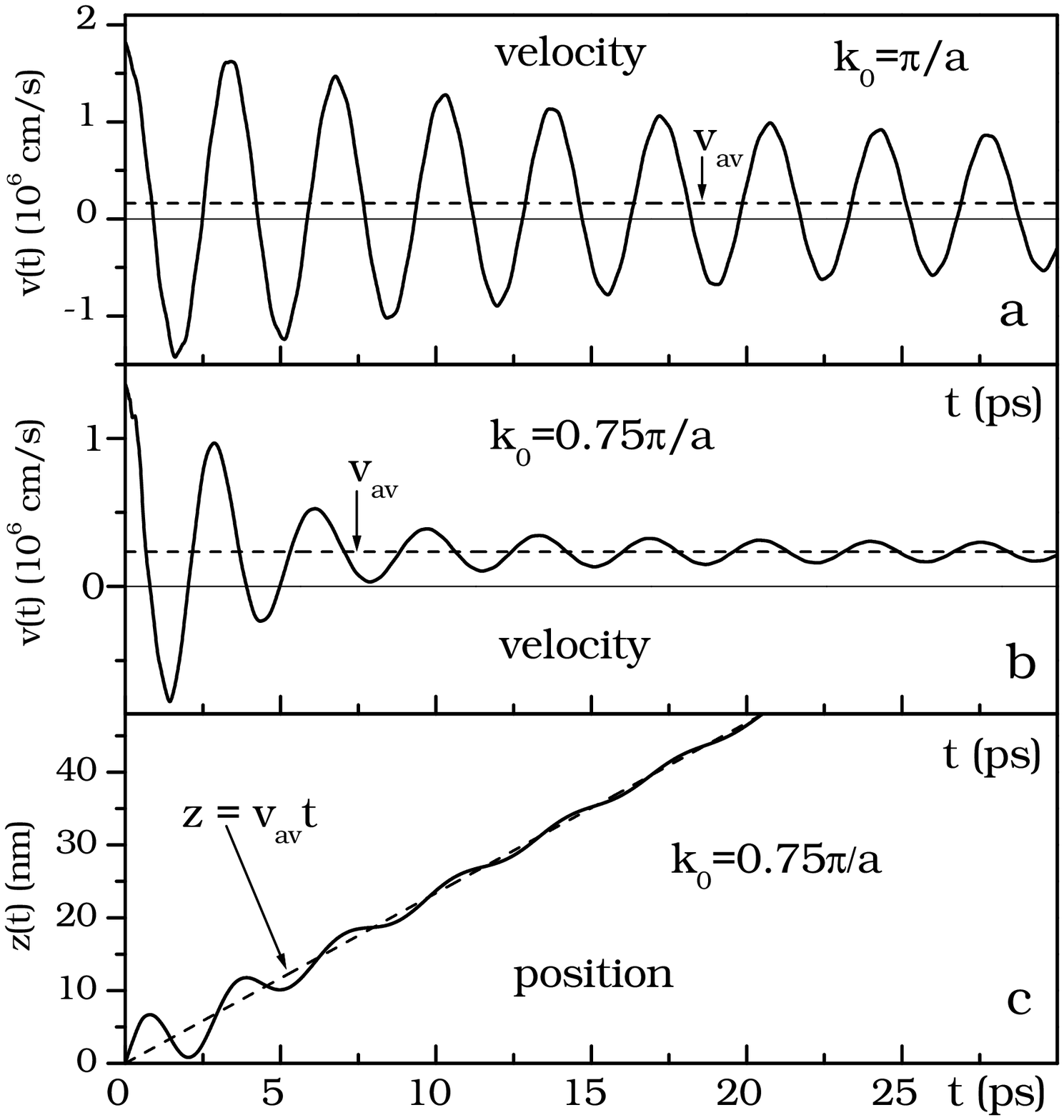}
\caption{Calculated electron ZB velocities and displacement in a superlattice versus time.
         The packet width is $d=400$\AA, Kronig-Penney parameter is $Z=1.5\pi$,
         superlattice period is $a=200$\AA. (a) Packet centered at $k_0=\pi/a$;
         (b) and (c) packet centered at $k_0=0.75\pi/a$.
         The dashed lines indicate motions with average velocities. After~\cite{Zawadzki2010}.}
\label{KronPFig2}
\end{figure}

Now we describe ZB using a rigorous quantum approach. We employ the Kronig-Penney
delta-like potential since it allows one to calculate explicitly the eigenenergies and
eigenfunctions~\cite{Kronig1931,SmithBook}.
In the extended zone scheme the Bloch functions are $\psi_k(z)=e^{ikz}A_k(z)$, where
\begin{equation} \label{Kron_Ak}
 A_k(z) = e^{-ikz}C_k\left\{e^{ika}\sin[\beta_kz] + \sin[\beta_k(a-z)]\right\},
\end{equation}
in which~$k$ is the wave vector, $C_k$ is a normalizing constant and $\beta_k=\sqrt{2m_0E}/\hbar$ is a
solution of the equation
\begin{equation}
 Z\frac{\sin(\beta_ka)}{\beta_ka} + \cos(\beta_ka) = \cos(ka),
\end{equation}
with $Z>0$ being an effective strength of the potential. In the extended zone scheme,
the energies~$E(k)$ are discontinuous functions for $k=n\pi/a$,
where $n=\ldots-1,0,1\ldots$.
In the Heisenberg picture the time-dependent velocity averaged over a wave packet~$f(z)$ is
\begin{equation} \label{Kron_v0}
\langle \hat{v}(t)\rangle = \frac{\hbar}{m_0} \int \hspace*{-0.5em}
        \int \!\!dk dk' \langle f|k\rangle\langle k|\frac{\partial}{i\partial z}|k'\rangle
                  \langle k'|f\rangle e^{i(E_k-E_{k'})t/\hbar},
\end{equation}
where $|k\rangle$ are the Bloch states. The matrix elements of momentum
$\langle k|\hat{p}|k'\rangle=\hbar \delta_{k',k+k_n}K(k,k')$ are calculated explicitly.
The wave packet~$f(z)$ is taken in a Gaussian form of the width $d$ and centered at~$k_0$.
Figure~\ref{KronPFig2} shows results for the electron ZB, as computed for a superlattice.
The electron velocity and position are indicated.
It is seen that for a superlattice with the period $a=200$\AA\ the ZB displacement is
about $\pm 50$\AA, i.e. a fraction of the period, in
agrement with the rough estimations given above. The period of oscillations is of the order of several
picoseconds.

\begin{figure}
\includegraphics[width=8cm,height=8cm]{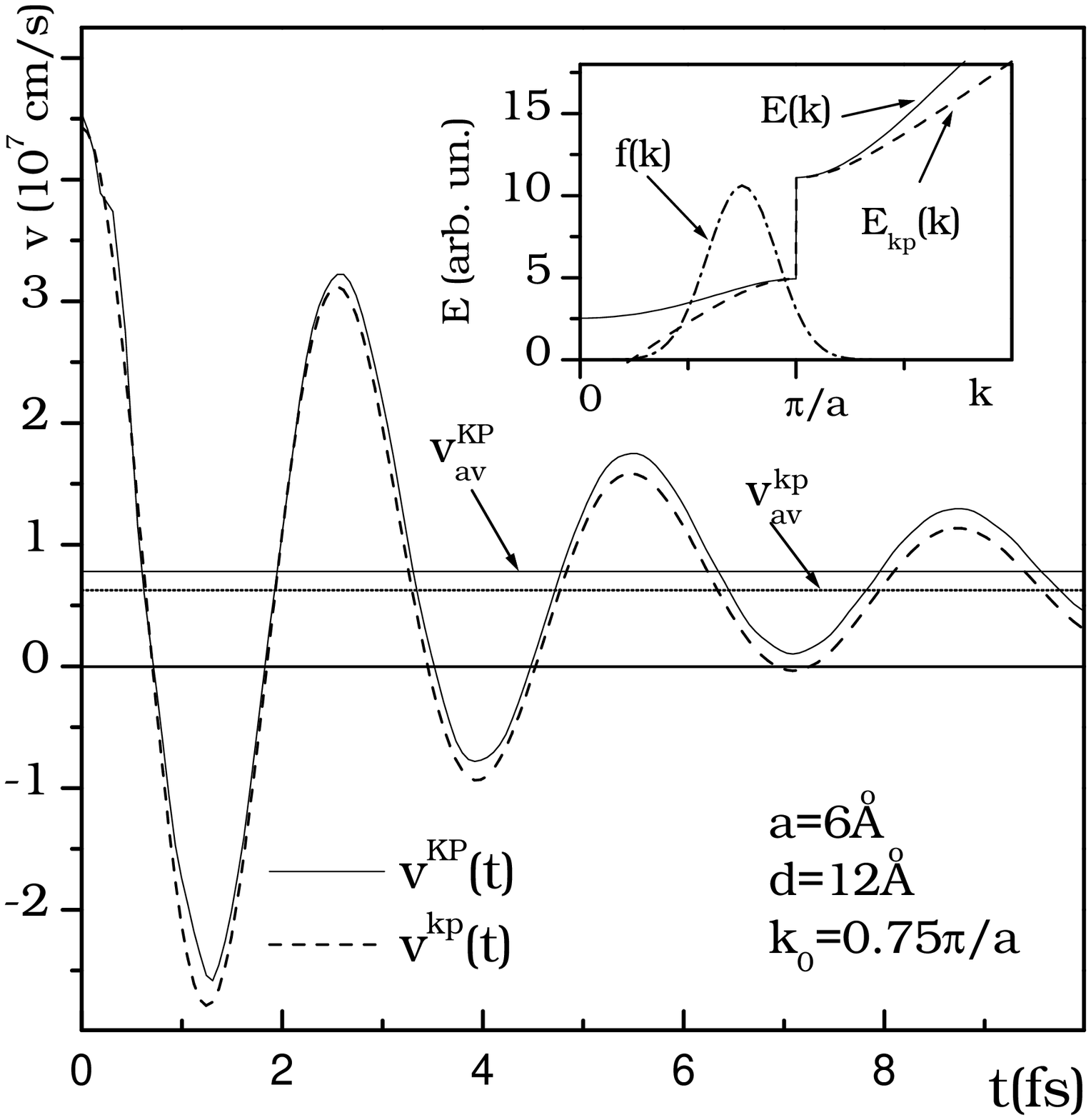}
\caption{Zitterbewegung of electron velocity in a periodic lattice versus time. Solid line: the Kronig-Penney
         model, dashed line: the two-band {\bf k.p} model. Inset: Calculated bands for the Kronig-Penney
         (solid line) and the two-level {\bf k.p} model (dashed line) in the vicinity of $k=\pi/a$.
         The wave packet~$f(k)$ centered at $k_0=0.75\pi/a$ is also indicated (not normalized).
         After~\cite{Zawadzki2010}. }
\label{KronPFig3}
\end{figure}

The oscillations of the packet velocity calculated directly from the periodic potential have many
similarities to those computed on the basis of the two-band {\bf k.p} model.
The question arises: does one deal with {\it the same} phenomenon in the two cases? To answer this
question we calculate ZB using the two methods for the same periodic potential. We calculate the packet
velocity near the point $k_0=\pi/a$ for a one-dimensional Kronig-Penney periodic Hamiltonian using
the Luttinger-Kohn (LK) representation~\cite{Luttinger1955}. The LK functions $\chi_{nk}(z)=e^{ikz}u_{nk_0}(z)$,
where $u_{nk_0}(z)=u_{nk_0}(z+a)$, also form a complete set and we can calculate the velocity
using a formula similar to~(\ref{Kron_v0}). The two-band model is derived by the {\bf k.p} theory
with the result
\begin{equation} \label{kp_Hkp}
 \hat{H}_{kp}= \left(\begin{array}{cc} \hbar^2q^2/2m + E_1 & \hbar qP_{12}/m \\
       \hbar qP_{21}/m & \hbar^2q^2/2m + E_2 \end{array}\right),
 \end{equation}
where~$E_1$ and~$E_2$ are the energies at band extremes,
$P_{12}=\hbar/m\langle u_{1k_{0}}|\partial/i\partial x|u_{2k_{0}}\rangle$, and $q=k-\pi/a$.
The band gap $E_q=E_2-E_1$ and the matrix elements $P_{12}$ are calculated from the
same Kronig-Penney potential, see inset of Figure~\ref{KronPFig3}.
Apart from the small free-electron terms on the diagonal, equation~(\ref{kp_Hkp}) simulates the 1+1
Dirac equation for free relativistic electrons in a vacuum.

In Figure~\ref{KronPFig3} we compare the ZB oscillations of velocity
calculated using: (a) real~$E(k)$ dispersions
resulting from the Kronig-Penney model and the corresponding Bloch functions of~(\ref{Kron_Ak});
(b) two-band~$E(k)$ dispersions and the corresponding LK functions. It is
seen that the two-band {\bf k.p} model
gives an excellent description of ZB for instantaneous velocities.
This agreement demonstrates that the theories based on: (a) the periodic potential and (b) the
band structure, describe {\it the same} trembling motion of the electron.
The procedure based on the energy band structure
is more universal since it also includes cases like the Rashba-type
spin subbands or the Luttinger-type light and heavy hole subbands which do not
exhibit an energy gap and do not seem to have a direct classical interpretation.
The distinctive character of the situation we considered is that it has a direct
spatial interpretation and it is in analogy to the situation first
considered by Schrodinger for a vacuum.

The main conclusion of the above considerations is that the electron
Zitterbewegung in crystalline solids is not an
obscure and marginal phenomenon but the basic way of electron propagation in a periodic potential.
The ZB oscillations of electron velocity are simply due to the total energy conservation.
The trembling motion can be described either as a mode of propagation in a periodic potential or,
equivalently, by the two-band {\bf k.p} model of band structure.
The latter gives very good results because, using the effective mass and the energy gap, it reproduces
the main features of the periodic potential.
According to the two-band model, the ZB is related to the interference of positive and negative
energy components, while the direct periodic potential approach reflects real character of this motion.
The established nature of ZB indicates that the latter should certainly be observable.

It should be mentioned that in their early paper Ferrari and Russo~\cite{Ferrari1990}
wrote: ``The motion of Zitterbewegung and the resulting formalism~\ldots is applied to describe the
acceleration of a non-relativistic electron moving in a crystal, due to the periodic force
experienced (\ldots). The resulting Zitterbewegung is a real effect just because it follows
from a real force.''

\section{Transport}

In this section we mention papers that relate the ZB phenomenon to the calculations of electron
transport in semiconductors.

Katsnelson~\cite{Katsnelson2006} used the Kubo and Landauer formalism to explain observed finite
minimum of the zero-temperature conductivity of monolayer and bilayer graphene at the vanishing
carrier density. He showed that it is the Zitterbewegung (interband) term in the current that
is responsible for this unusual behavior of conductivity in such extreme conditions.

Trauzettel {\it et al.}~\cite{Trauzettel2007} discussed photon-assisted electron transport in ballistic graphene
related to electron ZB in this material and concluded that, while the considered setup is
potentially relevant to the detection of ZB, the fundamental signature of ZB needs more
precise identification.

Cserti and David~\cite{Cserti2010} showed recently that the charge conductivity of the impurity-free
conductor can be expressed by non-diagonal amplitudes of ZB, while the Berry curvature and the Chern
number are related to the diagonal ZB parts. The developed method was applied to calculate
electric conductivity of various systems.

\section{Relativistic electrons in a vacuum}
The subject of ZB for free relativistic electrons is vast and we can not possibly do justice to it.
We mention below a few papers which contributed to the understanding of ZB in solids and its
simulations in other systems. The main idea of Schrodinger's pioneering work is given in~(\ref{z(t)})
because the initial equation~(\ref{H_WZ}) is the same as the Dirac equation (DE)
with changed parameters. Details of
the original Schrodinger derivations were given by Barut and Bracken~\cite{Schroedinger1930}.
Considerations showing that the ZB is caused by the interference of electron states related to positive and
negative electron energies are quoted in most books on relativistic quantum mechanics,
see e.g.~\cite{BjorkenBook,GreinerBook,SakuraiBook}. Feschbach and Villars~\cite{Feshbach1958} argued that,
in addition to the so-called Darwin term, the spin-orbit term in the standard~$v^2/c^2$ expansion
of DE can also be related to ZB.

Huang~\cite{Huang1952} went beyond the operator considerations of ZB calculating averages
of the electron position and angular momentum with the
use of wave packets. According to this treatment the electron magnetic moment may be viewed
as a result of ZB, see also~\cite{Tani1951}.
Huang did not predict the transient character of ZB since he assumed a very narrow
packet in~$k$ space (see~\cite{Lock1979}). Foldy and Wouthuysen~\cite{Foldy1950}
(see also~(\cite{Pryce1948,Tani1951}))
found a unitary transformation that separates the states of positive and negative electron
energies in the free-electron DE. They showed that such states do not exhibit the ZB.

Lock remarked that, in order to talk seriously about observing ZB, one should consider a localized electron
since ``it seems to be of limited practicality to speak of rapid fluctuations in the average
position of a wave of infinite extent''. He then showed that, if an electron is represented by a
localized wave packet, its Zitterbewegung is transient, i.e. it decays in time. This prediction was
subsequently confirmed by many descriptions (beginning with~\cite{Rusin2007b,Thaller2004} and
in experimental simulation~\cite{Gerritsma2010}. Lock further showed that, if the electron spectrum is discrete,
the resulting ZB is sustained in time. This property was confirmed for graphene in the presence of
an external magnetic field~\cite{Rusin2008}, as well as for relativistic electrons in a vacuum
when the spectrum is quantized into Landau levels~\cite{Rusin2010}.

It was pointed out, see e.g.~\cite{SakuraiBook,ThallerBook}, that according to the DE not only
the velocity and position operators experience ZB, but also the angular momentum $\hat{\bm L}$,
the spin $\hat{\bm S}$, and the operator $\hat{\beta}$ exhibit the trembling time dependence.
On the other hand, the total angular momentum $\hat{\bm J} = \hat{\bm L} + \hat{\bm S}$
is a constant of the motion, which can be shown directly by its vanishing commutator with the
free electron Hamiltonian $\hat{\cal H}_D$.

Braun {\it et al.}~\cite{Braun1999} used a split-operator technique to solve
numerically the 3D time-dependent DE.
Gaussian wave packets we employed to calculate the transient ZB in the position and spin of free
relativistic electrons for different packet widths. The authors remark that ``the Zitterbewegung can be
found only if the initial velocity (or wave vector)~$k_0$ is nonzero''. Very good numerical approximations
to the exact solutions were found but they require powerful computers.

Thaller~\cite{Thaller2004} computed and simulated the time behavior of relativistic Gaussian wave packets
according to the one-dimensional DE. For a packet with vanishing average momentum, the
packet position shows ZB that decays with time very slowly. For a non-vanishing average
momentum the decaying of ZB is much faster. This is caused by the fact that the ZB arises due to
an interference of positive and negative energy sub-packets which in this case move in
opposite directions and cease to overlap relatively quickly. This process is explained in
some detail in~(\ref{BG_psip}) and~(\ref{BG_psin}) for electrons in bilayer graphene.

Krekora {\it et al.}~\cite{Krekora2004} studied pair creation in a vacuum and stated that ``quantum
theory prohibits the occurrence of Zitterbewegung for an electron''. This conclusion was contradicted
by the analysis of Wang and Xiong~\cite{Wang2008}.
Arunagiri~\cite{Arunagiri2009} proposed to circumvent the difficulty preventing observability
of ZB due to pair creation by localizing the electron in the presence of a magnetic field and
using monolayer graphene as a model of massless fermions.
Barut and Malin~\cite{Barut1968} considered
the problem of filled negative energies in the Dirac equation and its effect on electron localization.

Barut and Thacker~\cite{Barut1985} treated the ZB of relativistic electrons in a vacuum in
the presence of an external magnetic field. This description suffered from a few
deficiencies, as explained in~\cite{Rusin2010}. Bermudez {\it et al.}~\cite{Bermudez2007} treated the
problem of time dependent relativistic
Landau states by mapping the relativistic model of electrons in a magnetic
field onto a combination of the Jaynes-Cummings and anti-Jaynes-Cummings
interactions. For simplicity the $p_z=0$
restriction was assumed. Three regimes of high (macroscopic), small
(microscopic) and intermediate (mesoscopic) Landau quantum numbers $n$ were
considered. In all the cases only one interband frequency contributed to the
Zitterbewegung because the authors did not use wave packets to calculate average values.
The same problem was recently tackled by Rusin and Zawadzki~\cite{Rusin2010} who showed that the
quantization of electron spectrum into the Landau levels has strong effects on the ZB.
The trembling motion becomes a multi-frequency phenomenon and in two dimensions is not transient,
as opposed to the no-field case. In practice, however, for magnetic fields available in terrestrial conditions
the decisive ratio $\hbar(eB/m_0)/2m_0c^2$ is very small, so the magnetic effects in the ZB are insignificant.
The only promising way to see the magnetic effects in ZB is to carry out simulations.
Such a simulation was proposed in~\cite{Rusin2010}, see section VIII.

\section{Simulations}
As we said above, the electron Zitterbewegung in a vacuum or in a solid is difficult to observe.
The characteristics of electron ZB in semiconductors are much more favorable than in a vacuum
but it is difficult to follow the motion of a single electron; one would need to follow motion of many
electrons moving in phase. Recently, however, there appeared many propositions to {\it simulate}
the Dirac equation and the resulting phenomena with the use of other systems. We want to enumerate
below these propositions but we are not in a position to explain all the underlying ideas. It will suffice to
say that many (not all) ideas make use of trapped atoms or ions interacting with laser light. There are
two essential advantages of such simulations. First, it is possible to follow the interaction of laser light
with few or even single atoms or ions. Second, when simulating the DE it is possible to modify its two
basic parameters: $mc^2$ and $c$, in order to make the ZB frequency much lower and its amplitude much larger
than in a vacuum. In consequence, they become measurable with current experimental techniques.

As a matter of example we will briefly consider a simulation of DE
with the use of Jaynes-Cummings model~\cite{Jaynes1963}
known from the quantum and atomic optics, see References~\cite{Lamata2007,Leibfried2003,Johanning2009}.
The Dirac equation contains electron momenta, so the essential task is to simulate $\hat{p}_l$. The common
types of light interactions with ions and vibronic levels are used to that purpose:
a carrier interaction $\hH^{c}=\hbar\Omega(\sigma^+e^{i\phi_c}+\sigma^-e^{-i\phi_c})$,
the Jaynes-Cummings interaction
$\hH^{\phi_r}_{JC}= \hbar\eta\tilde{\Omega}(\sigma^+\ha e^{i\phi_r}+\sigma^-\hap e^{-i\phi_r})$,
and the anti-Jaynes-Cummings interaction
$\hH^{\phi_b}_{AJC}=\hbar\eta\tilde{\Omega}(\sigma^+\hap e^{i\phi_b}+\sigma^-\ha e^{-i\phi_b})$.
Here $\sigma^{\pm}=\sigma_x\pm i\sigma_y$ are the raising and lowering ionic spin-$1/2$ operators,
$\ha$ and $\hap$ are the creation and annihilation operators associated with the motional states of
the ion, $\eta$ is the so called Lamb-Dicke parameter, $\Omega$ and $\tilde{\Omega}$ are the Rabi
frequencies. The basic idea is to use proper light phases in $\hH^{\phi_r}_{JC}$ and $\hH^{\phi_r}_{AJC}$
in order to obtain the momentum from the relation $\hat{p}_l=i\hbar(\hap_l-\ha_l)/\Delta$, which
results in $\hH^{p_l}_{\sigma_j}=\pm i\hbar \eta \Delta \tilde{\Omega}\sigma_j(\hap_l-\ha_l)$.
One can show that the simulated parameters of DE are
\begin{equation}
c\rightarrow 2\eta\Delta\tilde{\Omega},\hspace*{2em} mc^2\rightarrow \hbar\Omega,
\end{equation}
where $\Delta$ is the spread in the position of the ground ion wave function. If the dynamics created by the
3+1 Dirac equation is to be reproduced in an experiment with a four-level ion system using Raman beams,
it requires 14 pairs of Raman lasers. One needs to control their phases independently.

\begin{figure}
\includegraphics[width=8.5cm,height=8.5cm]{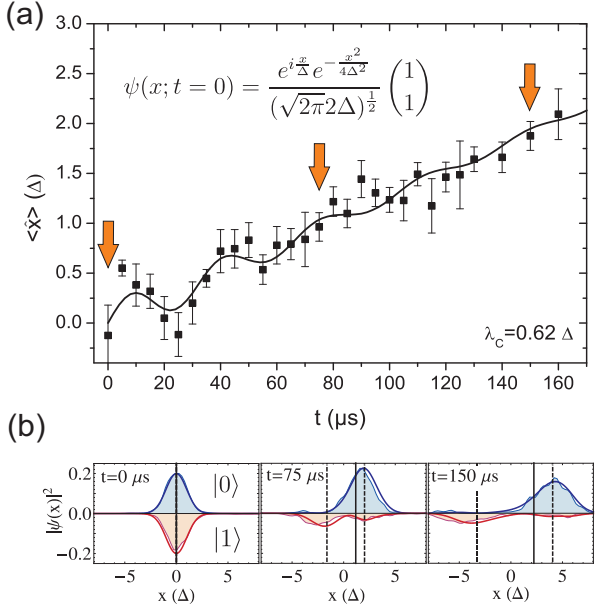}
\caption{ (a)Zitterbewegung for a state with non-zero average momentum.
          The solid curve represents a numerical simulation. (b) Measured (filled areas) and numerically
          calculated (solid lines) probability distributions $|\psi(x)|^2$  at the
          times $t=0$,~75 and 150~$\mu$s (as indicated by the arrows in (a)).
          The probability distribution corresponding to the state $|1\rangle$ is inverted for clarity.
          The vertical solid line represents $\langle \hat{x} \rangle$ as plotted in (a).
          The two dashed lines are the expectation values for the positive and negative energy
          parts of the spinor. Error bars 1$\sigma$. After~\cite{Gerritsma2010}.} \label{GerritsmaFig2}
\end{figure}

\begin{figure}
\includegraphics[width=8.5cm,height=8.5cm]{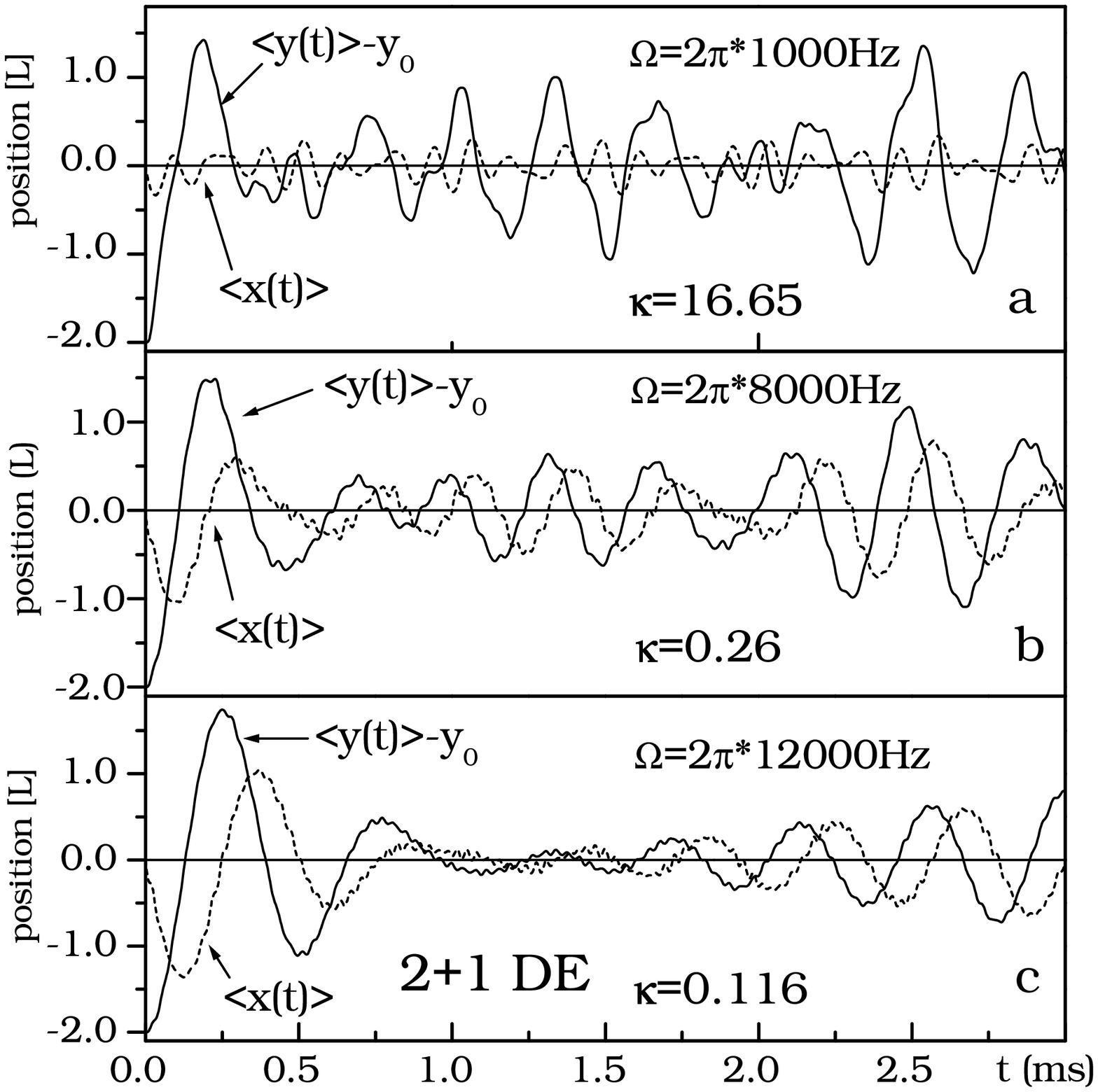}
\caption{Calculated motion of two-component wave packet simulated
         by trapped $^{40}$Ca$^{+}$ ions for three values of effective rest energies~$\hbar\Omega$.
         Simulations correspond to $\kappa=\hbar\omega_c/2mc^2$=
         16.65~(a), 0.26~(b), 0.116~(c), respectively. Positions are given in $L=\sqrt{2}\Delta$ units.
         Oscillations do not decay in time. After~\cite{Rusin2010}.} \label{SimFig1}
\end{figure}

In Figure~\ref{GerritsmaFig2} we show experimental results of Gerritsma {\it et al.}~\cite{Gerritsma2010},
who simulated for the first time the 1+1 Dirac equation with the resulting
one-dimensional Zitterbewegung using $^{40}$Ca$^{+}$ trapped ions.
It can be seen that the results agree very well with the predictions of~\cite{Zawadzki2010}, see our
Figure~\ref{KronPFig2}c. The reason of this agreement is that the
theory of~\cite{Zawadzki2010}, while concerned with solids, also uses an effective Dirac equation,
see~(\ref{kp_Hkp}) and the inset of Figure~\ref{KronPFig3}. Gerritsma {\it et al.} showed that,
if the wave packet does not have
the initial momentum, the decay time of ZB is much slower than that seen in Figure~\ref{GerritsmaFig2}.
This agrees with theoretical results for carbon nanotubes, as shown in our Figure~\ref{ZitterGrLFig3},
see also~\cite{Thaller2004}.

The problem of ZB for free relativistic electrons in a magnetic field was recently described by
Rusin and Zawadzki~\cite{Rusin2010}.
The main experimental problem in investigating the ZB phenomenon in an external magnetic field
is the fact that for free relativistic electrons the basic ZB (interband) frequency corresponds to
the energy $\hbar\omega_Z \simeq 1$~MeV, whereas the cyclotron frequency for a magnetic
field of 100~T is $\hbar\omega_c \simeq 0.1$~eV, so that the magnetic effects in ZB are
very small. However, it is possible to simulate the Dirac equation including an external
magnetic field with the use of trapped ions interacting with laser radiation.
This gives a possibility to modify the ratio $\hbar\omega_c/(2mc^2)$ making its value much more
advantageous. If the magnetic field $\bm B$ is directed along the $z$ direction, it can be described
by the vector potential $\bm A = (-By,0,0)$. Then the main modification introduced to the DE is that,
instead of the momentum $\hat{p}_x$, one has $\hat{\pi}_x=\hat{p_x}-eBy$
which leads to the appearance of raising
and lowering operators $\ha_y =(\xi+ \partial/\partial \xi )/\sqrt{2}$ and
$\hap_y= (\xi -\partial/\partial \xi)/\sqrt{2}$ with $\xi=y/L-k_xL$. The simulation of DE
proceeds as for a free particle with the difference that $\ha_y$ and $\hap_y$ can be simulated by
a {\it single} JC or AJC interaction. It can be shown that the crucial ratio is
\begin{equation} \label{IonKappa}
 \kappa = \frac{\hbar eB}{m(2mc^2)} \Rightarrow \left(\frac{\eta\tilde{\Omega}}{\Omega}\right)^2,
\end{equation}
where $\eta$, $\Omega$ and $\tilde{\Omega}$ were defined above.
Therefore, by adjusting frequencies $\Omega$ and $\tilde{\Omega}$ one can simulate different regimes of
$\kappa=\hbar\omega_c/2mc^2$. In Figure.~\ref{SimFig1} we show the calculated ZB for three values of~$\kappa$:
16.65, 1.05, 0.116. It is seen that, as~$\kappa$ gets larger (i.e. the field intensity increases or the
effective gap decreases), the frequency spectrum of ZB becomes richer.
This means that more interband and intraband frequency components contribute to the spectrum.
Both intraband and interband frequencies correspond to the selection rules $n'=n\pm 1$ so that,
for example, one deals with ZB (interband) energies between the Landau levels $n=0$ to $n'=1$, and
$n=1$ to $n'=0$, as the strongest contributions. For high magnetic fields the interband and intraband
components are comparable. Qualitatively, the results shown in Figure~\ref{SimFig1} are similar
to those obtained for graphene in a magnetic field, see~\cite{Rusin2008}.

Vaishnav and Clark~\cite{Clark2008} proposed to observe the ZB with ultra-cold neutral atoms in an
optical lattice. One can show that such a system in a tripod configuration simulates the
Dirac equation with the resulting ZB. The characteristic ZB amplitude is equal to the wavelength
of light producing the optical lattice and the light velocity is replaced by the lattice
recoil velocity of a few cm/s. The simulated ZB has accessible characteristics: the amplitude of
tens of nm and frequencies in the range of MHz. Interestingly, the transient ZB occurs also for a
vanishing average momentum of the wave packet. Also, the ZB can be viewed as a measurable
consequence of the momentum-space Berry phase.

Zhang {\it et al.}~\cite{Zhang2010} extended the idea of Vaishnav and Clark~\cite{Clark2008}
for the tripod scheme proposing that one can use vibrating mirrors to modulate the
laser light and simulate the time-dependent Dirac equation in order to influence
the amplitude, frequency and decay time of the resulting Zitterbewegung.

Merkl {\it et al.}~\cite{Merkl2008} described the ZB of ultra-cold atoms moving in one-dimension and
interacting with laser beams in a tripod system. It was shown explicitly that in this case the decay time of
ZB is inversely proportional to the~$k$ spread of a wave packet and that the oscillation amplitude
decreases as~$t^{-1/2}$. These features agree with the analytical results shown above for the ZB of
electrons in bilayer graphene, c.f.~(\ref{BG_x(t)})-(\ref{BG_xZ(t)}).

Song and Foreman~\cite{Song2009} proposed to create the atomic ZB using trapped cold atoms in an Abelian
vector potential in a tripod configuration. It was shown that, in a purely 1D potential, one can
still achieve time-dependent velocity operator if the scalar potential does not commute with the
vector potential. The predicted amplitude of transient ZB is around 0.2$\mu$m and the frequency
around 1 ms$^{-1}$.

Braun~\cite{Braun2010} showed how a single harmonically trapped cold atom in a real spatially
tailored magnetic field can be used to simulate the Bychkov-Rashba and the linear Dresselhaus
spin coupling with the resulting Zitterbewegung.

\section{ZB-like wave effects}

In its original version proposed by Schrodinger the Zitterbewegung is a quantum
phenomenon {\it par excellence} since it predicts the electron behavior that goes beyond Newton's
fist law of classical motion. However, a similar behavior is predicted and also observed in the
propagation of acoustic and light waves in periodic systems. These ZB-like effects result from the
wave nature of phonons and photons and they are in principle of a non-quantum nature (they do not involve
the Planck constant). Clearly, they do not go beyond Newton's first law of motion, similarly to the
electron ZB effects in solids which, as we demonstrated above, are related to the periodic potential of
the crystal lattice. We emphasize here that the acoustic and light effects mentioned below do not
simulate the relativistic quantum mechanics (as they are sometimes presented), but represent
the wave ZB-like effects in other systems.

Zhang and Liu~\cite{ZhangLiu2008} investigated experimentally acoustic wave propagation in 2D
macroscopic sonic crystals made of steel cylinders immersed in water and having a lattice
constant $a=1.5$mm. The band structure of such crystals for acoustic waves has no energy gap
and almost linear~$E(\bm q)$ dependence at the~$K$ point of the Brilloiun zone. As a consequence,
the acoustic wave propagation near the~$K$ point can be described by a $2\times 2$ set of equations resembling
the Dirac-like Hamiltonian for monolayer graphene. The observed effects in the experimental behavior
of transmitted acoustic waves were interpreted as a classical analogue of ZB.

Wang {\it et al.}~\cite{Wang2010} described sound propagation in
sonic crystals consisting of square arrays
of steel cylinders immersed in water. For acoustic waves, the band structure $\omega(q)$ of such systems
resembles the relativistic two-band dispersion. It is predicted that
the time evolution of an acoustic wave packet
(pressure intensity) should exhibit a transient ZB-like effect with the initial amplitude of about one
lattice constant and frequency $\omega_Z=\omega_2(q_0)-\omega_1(q_0)$, equal to the frequency
difference of the bands in question. The transient character of ZB is due to a weakening interference
of subpackets as they move apart. All these features resemble very closely the electron ZB in crystalline
solids.

Zhang~\cite{Zhang2008PRL} simulated numerically the photon transport in 2D photonic crystals
made of cylinders immersed in air. In such a crystal, the band structure for photons near the~$K$ point
of the Brillouin zone can be described by the Dirac equation with zero gap and no spin. If a photon
wave packet is propagating through the crystal, the light intensity at various points, both inside and
outside of it, exhibits the ZB behavior.
The photon propagation in photonic crystals bears many similarities to the sound
propagation in sonic crystals, see~\cite{ZhangLiu2008}.

Wang {\it et al.}~\cite{Wang2009} used the fact that in a homogeneous optical medium consisting of
three slabs characterized by negative-zero-positive refractive indices there exist two optical
pass-bands for photons with linear~$E(k)$ dispersions, crossing at the so-called Dirac point.
If one sends an optical pulse with frequencies near the Dirac point (GHz region), the pulse
exhibits ZB-like oscillations due to the interference of states in the upper and lower
high-transmittance bands.

Longhi~\cite{Longhi2010a} showed that in nonlinear optics (sum frequency generation) one also deals with
processes that mimic the one-dimensional Dirac equation. As a consequence, he predicted the Zitterbewegung
of short optical pulses in nonlinear quadratic media.

Dreisow {\it et al.}~\cite{Dreisow2010}, following the suggestion of Longhi~\cite{Longhi2010b},
realized lately an optical binary waveguide system which was shown to have for photons the relativistic-like
dispersion $E(k)=\sqrt{\sigma^2+\kappa^2k^2}$. The energy gap is $2\hbar\sigma$, where
$\sigma$ is the mismatch of propagation constants and $\kappa$ is the coupling rate between two
adjacent waveguides. The resulting trembling motion was observed as a spatial oscillatory motion
of an optical beam with the frequency $\omega_Z=2\sigma$ and amplitude $R_Z=\kappa/(2\sigma)$. The
experiments were carried out for highly relativistic (small $\sigma$) and weakly relativistic
(higher $\sigma$) regimes.

The ZB-like wave phenomena in periodic structures are very similar to the electronic ZB in
crystalline solids: they are characterized by the interband frequency, they result from an interference
of states related to the positive and negative energies and they decay in time. Paradoxically,
these wave phenomena seem to be easier to observe than their ``older'' electronic analogues.

\section{Discussion and conclusions}
An important recognition won after the considerable effort of the last five years is, that the Zitterbewegung
is not a marginal, obscure and probably unobservable effect of interest to a few esoteric theorists,
but a real and universal phenomenon that often occurs in both quantum and non-quantum systems.
Clearly, the ZB in a vacuum proposed by Schrodinger~\cite{Schroedinger1930} stands out as an exception since
it is supposed to occur without any external force. However, in its original form it is probably not
directly observable for year to come and one has to recourse to its simulations. A proof-of-principle of
such simulations was recently carried out, see Figure~\ref{GerritsmaFig2} and Reference~\cite{Gerritsma2010}.
On the other hand, manifestations of ZB in crystalline solids and other periodic systems turned out to
be quite common and they are certainly observable. A universal background underlying the phenomenon of ZB
in any system (including a vacuum) is an interference of states belonging to positive and negative energies
(in a generalized sense, see below). The positive and negative energies belong usually to
bands but they can also be discrete levels, as shown for electrons in
graphene in a magnetic field, see Figure~\ref{GraphHFig1}.

As we said above, the ZB amplitude is around $\lambda_Z=\hbar/(m_0^*u)$, which we called the length
of Zitterbewegung. Let us suppose that we confine an electron to the dimension $\Delta z \simeq \lambda_Z/2$.
Then the uncertainty of momentum is $\Delta p_z\ge \hbar/\Delta z$ and the resulting uncertainty
of energy $\Delta E\simeq (\Delta p_z)^2/(2m_0^*)$
becomes $\Delta E \ge 2m_0^* u^2={\cal E}_g$. Thus the
electron confined to $\Delta z \simeq \lambda_Z/2$ has the uncertainty of energy larger than the gap.
For electrons in a vacuum the restriction $\Delta z \simeq \lambda_c/2$ is not significant, but for
electrons in narrow-gap semiconductors the restriction $\Delta z \simeq \lambda_Z/2$ should be taken
seriously since $\lambda_Z$ is of the order of tens of Angstroms, so that this confinement is not difficult
to realize experimentally by quantum wells or magnetic fields. The question arises, what happens
if the electron is confined to $\Delta z < \lambda_Z/2$, so that the trembling motion is strongly
perturbed by the confinement.
We showed above, see Figure~\ref{GraphHFig6}, that an electron in a magnetic field
radiates interband ZB frequencies and their contribution to the motion increases with the increasing field,
see also~\cite{Rusin2010}. It is possible that this effect is just a manifestation of the perturbation
of the trembling motion by magnetic confinement. Also, it was shown that an effective one-band
semi-relativistic Hamiltonian in a narrow-gap semiconductor contains the so-called Darwin term which
can be traced back to the ZB. The Darwin term can lead to measurable effects for ground
impurity states~\cite{Zawadzki2005KP}.

An important question arises: what should be called ``Zitterbewegung''?
It seems that the signature of ZB phenomenon is its {\it interband frequency}, in which the term
{\it interband} has the meaning ``between interacting bands''. Thus, for example, the ZB resulting from
the Bychkov-Rashba spin splitting (or the so-called linear Dresselhaus spin splitting) is not
characterized by a truly interband frequency, since in this situation there is no gap, but the
frequency corresponding to the energy difference between the two spin branches of the
same band: $\hbar\omega_Z=E_{\uparrow}-E_{\downarrow}$, see~(\ref{Sch_x(t)})
and Reference~\cite{Schliemann2005}.
Another illustration is the ZB of holes in the valence bands of $\Gamma_8$ symmetry~\cite{Luttinger1956},
where the ZB frequency is given by the energy difference of light and heavy hole
bands~\cite{Jiang2005,Winkler2007,Demikhovskii2010}. Finally, an instructive example is provided
by graphene in a magnetic field (see Figure~\ref{GraphHFig1}), where the electron motion contains both
intraband and interband frequencies. We believe that only the interband contributions should be
called ZB, while the intraband ones are simply the cyclotron components. It appears that the second
signature of ZB is the actual {\it motion} which, for instance, distinguishes it from the Rabi oscillations.
The above considerations indicate that an unambiguous definition of ZB is not obvious.

If an electron is prepared in the form of a wave packet, and if the electron spectrum is not completely
quantized, the ZB has a {\it transient} character, i.e. it decays in time.
This was predicted by Lock~\cite{Lock1979}
on the basis of the Riemann-Lebesgue lemma, and was confirmed by many specific calculations, see e.g.
Figures~\ref{ZitterGrLFig1} and~\ref{ZitterGrLFig2}, as well
as by observations~\cite{Gerritsma2010,ZhangLiu2008}. One can show that
the decay time is inversely proportional to the momentum spread $\Delta k$ of the wave packet,
see~(\ref{BG_GammaZ}) and References~\cite{Rusin2007b,Merkl2008}.
Physically, the transient character of ZB comes about as a result of
waning interference of the two sub-packets belonging to positive and negative energies as they go apart
because of different speeds, see~(\ref{BG_psip}) and~(\ref{BG_psin}) and References~\cite{Rusin2007b,Wang2010}.
The decay time is usually much longer in one-dimensional systems, see Figure~\ref{ZitterGrLFig3} and
References~\cite{Rusin2007b,Merkl2008}.
On the other hand, if the electron spectrum is discrete, ZB persists in time, sometimes in the form of
collapse-revival patterns~\cite{Romera2009}.
In general, the wave packet should have a non-vanishing initial momentum in one direction to exhibit the ZB
in the perpendicular direction see~(\ref{Sch_x(t)}) and~(\ref{BG_x_11}),
but this is not always the case~(\cite{Maksimova2008}).

We described above the phenomena related to photons and phonons in separate Section IX because, in our
opinion, they are not simulations of the relativistic quantum mechanics but represent ZB-like effects
of their own. They are non-quantum wave effects in periodic structures. With the quantum electronic
ZB effects they have in common the Floquet and Mathieu descriptions of eigenvalues and eigenfunctions
of the second-order differential equations with periodic potentials, see e.g.~\cite{Brillouin1956}.
Quantum and non-quantum phenomena can be quite similar because of the wave character of quantum
mechanics. It is an important success of the efforts concerned with the electronic ZB that it has lead to
the discoveries in non-electronic areas. It appears that, in fact, the non-electronic ZB-like effects
are easier to observe than the ``original'' electronic ones. Finally, it is important that photons and
phonons in the ZB-like wave phenomena do not obey the Pauli exclusion principle for fermions.

Clearly, one should ask the question about possible observation of Zitterbewegung in solids. Two
different ways were proposed to observe the trembling electrons. The first is to detect an ac current
related to the ZB velocity, see e.g. Figure~\ref{ZitterGrLFig2}.
One needs a current meter sensitive to the ZB frequency.
Then, even if the electrons do not move in phase so that the net current averages to zero, the meter
should detect a clear increase of noise at the frequency~$\omega_Z$. The second possible way to observe
the ZB is to detect electromagnetic radiation emitted by the trembling electrons,
see Figure~\ref{GraphHFig6} and Reference~\cite{Rusin2008}. The emission is possible because, if the
electrons are prepared in form of wave packets or they respond to light wave packets, they are
not in their eigenstates. The proposed ZB should not be confused with the
Bloch oscillations of charge carriers in superlattices.
The Bloch oscillations are basically a one-band phenomenon and they require an external
electric field driving electrons all the way to the Brillouin zone boundary.
On the other hand, the ZB needs at least two bands
and it is a no-field phenomenon. Narrow gap superlattices can provide a suitable system for its
observation. In the near future one can expect theoretical predictions of ZB in new systems as well
as observations of ZB-like wave effects. A real challenge remains: a direct experimental evidence
for the electron Zitterbewegung in semiconductors.

\appendix
\section{}

\begin{figure}
\includegraphics[width=8.5cm,height=8.5cm]{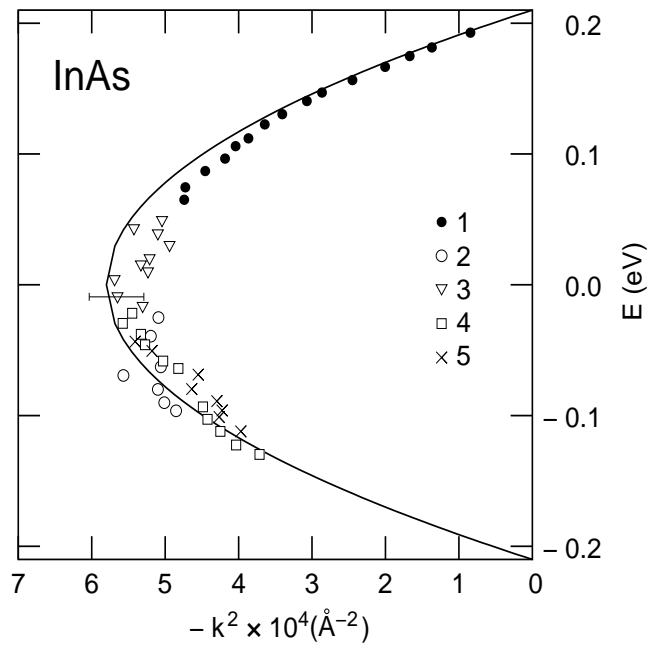}
\caption{\label{ZawadzkiFig1}Energy-wave vector dependence in the
forbidden gap of InAs. Various symbols show experimental data of
Parker and Mead~\cite{Parker1968} for five InAs samples, the solid line is
theoretical fit using~(\ref{A2Bands}). The determined parameters are
$\lambda_Z$ = 41.5 {\rm \AA} and $u$ = 1.33$\times 10^8$ cm/sec. After~\cite{Zawadzki2005KP}.}
\end{figure}

In this Appendix we show that the ZB length $\lambda_Z$ defined in~(\ref{E_LambdaZ})
can be measured directly. We write~(\ref{E_2bands}) in the form~\cite{Zawadzki2005KP}
\begin{equation} \label{A2Bands}
E=\pm\hbar u \left(\lambda^{-2}_Z + k^2\right)^{1/2}\;\;.
\end{equation}
where $u=(E_g/2m_0^*)^{1/2}$.
For $k^2 > 0$ this formula describes the conduction and
light-hole bands. But for imaginary values of~$k$ there is $k^2 <
0$ and~(\ref{A2Bands}) describes the dispersion in the energy gap. This
region is classically forbidden but it can become accessible
through quantum tunneling. Figure~\ref{ZawadzkiFig1} shows the data for the
dispersion in the gap of InAs, obtained by Parker and Mead~\cite{Parker1968}
from tunneling experiments with double Schottky barriers. The
solid line indicates the fit using~(\ref{A2Bands}). The value of
$\lambda_Z$ is determined directly by $k_0$ for which the energy
is zero: $\lambda^{-2}_Z = k^2_0$. The fit gives $\lambda_Z
\approx$ 41.5 {\rm \AA} and $u \approx 1.33\times 10^8$ cm/sec,
in good agreement with the estimation for InAs given in Section II.
Similar data for GaAs give $\lambda_Z$ between 10 {\rm \AA}~\cite{Padovani1966} and
13 {\rm \AA}~\cite{Conley1967}, again in good agreement with the estimation quoted in Section II.

\section{}
We briefly discuss here the classical electron velocity and mass for a linear energy band
of monolayer graphene, as they are often subjects of misunderstandings. Let us consider
the conduction band and take $p\ge 0$, where the pseudo-momentum is ${\bm p}=\hbar {\bm k}$. Then the
band dispersion is $E=up$ and the classical velocity is
\begin{equation} \label{B1}
v_i = \frac{\partial E}{\partial p_i} =\frac{dE}{dp} \frac{\partial p}{\partial p_i} =
    \frac{dE}{dp} \frac{p_i}{p} = \frac{dE}{dp} \frac{1}{p} \delta_{ij}p_j,
\end{equation}
where $\delta_{ij}$ is the Kronecker delta function and we use the sum convention over the
repeated index $j=1,2$. The electron mass tensor $\hat{m}$ relating the velocity to
momentum is {\it defined} by $\hat{m}{\bm v}={\bm p}$. Then the inverse mass tensor~$\hat{(1/m)}$
is defined by
\begin{equation} \label{B2}
v_i = \left(\frac{1}{m}\right)_{ij} p_j.
\end{equation}
Equating~(\ref{B1}) with~(\ref{B2}) we obtain
\begin{equation} \label{B3}
\left(\frac{1}{m}\right)_{ij} = \frac{dE}{dp} \frac{1}{p} \delta_{ij}.
\end{equation}
Thus the inverse mass tensor is a scalar: $1/m=(dE/dp)(1/p)$. Using the initial band dispersion
one has $dE/dp=u$, so that $m=p/u=E/u^2$. This equality can be seen in two ways. First, it gives
\begin{equation} E=mu^2, \end{equation}
which is analogous to the Einstein relation between the particle energy and mass. Second, the
formula
\begin{equation} m=\frac{E}{u^2} \end{equation}
states that the mass vanishes at $E=0$ (or $p=0$), but it is nonzero for $E>0$ (or $p>0$). These
relations hold also for a more general ``semi-relativistic'' case of narrow-gap semiconductors
described by~(\ref{E_2bands}), see References~\cite{ZawadzkiHMF,Zawadzki2006}.

One should add that, if one defined the mass by the relation of the force to
acceleration: $\hat{M}{\bm a}={\bm F}$, the inverse mass would be given by the {\it second derivative}
of the energy with respect to momentum. For the linear band of graphene:~$E=up$,
such a mass would be infinitely large for all energies, so it is not a useful quantity.

\hspace*{1em}

\end{document}